\documentclass[journal]{IEEEtran}
\IEEEoverridecommandlockouts
\usepackage[linesnumbered, ruled]{algorithm2e}

\usepackage{booktabs}
\usepackage{cite}
\usepackage{amsmath,amssymb,amsfonts}
\usepackage{algorithmic}
\usepackage{multirow}
\usepackage{makecell}
\usepackage{graphicx}
\usepackage{textcomp}
\usepackage{xcolor}
\usepackage{stfloats}
\usepackage{array}
\usepackage{bm}
\usepackage{epstopdf}
\usepackage{enumerate}
\usepackage{subcaption}
\usepackage{caption,setspace}
\begin{document}
	\title
	{
		 \huge A General Non-Stationary Channel Model for Autonomous Vehicle Communications in Underground Spaces
	}
	\author
	{
		Lin Hou, \textit{Graduate Student Member, IEEE}, Cheng-Xiang Wang, \textit{Fellow, IEEE}, Songjiang Yang, \textit{Member, IEEE}, \\Hengtai Chang, \textit{Member, IEEE}, Jie Huang, \textit{Senior Member, IEEE}, \\and El-Hadi M. Aggoune,  \textit{Life Senior Member, IEEE}
		\thanks{Copyright (c) 2026 IEEE. Personal use of this material
			is permitted. However, permission to use this material for any
			other purposes must be obtained from the IEEE by sending a
			request to pubs-permissions@ieee.org.}
		\thanks{This work was supported by the Major Science and Technology Project of Jiangsu Province under Grant BG2025039, the National Natural Science Foundation of China (NSFC) under Grant 62401644, the Research Fund of National Mobile Communications Research Laboratory, Southeast University, under Grant 2026A05, the Shandong Provincial Natural Science Foundation under Grant ZR2025MS1096, and the Promising Research Program, University of Tabuk, Saudi Arabia, under Grant PRP-2025-01.
		\textit{(Corresponding Authors: Cheng-Xiang Wang; Songjiang Yang)}.
		}
		\thanks{L. Hou, C.-X. Wang, and J. Huang are with the National Mobile Communications Research Laboratory, School of Information Science and Engineering, Southeast University, Nanjing 211189, China, and also with Purple Mountain Laboratories, Nanjing 211111, China (e-mail: \{hou\_lin, chxwang, j\_huang\}@seu.edu.cn).}
		\thanks{S. Yang is with the School of Communication and Information Engineering, Shanghai University, Shanghai 200444, China (e-mail: yangsongjiang@shu.edu.cn).}
		\thanks{H. Chang is with the Shandong Provincial Key Laboratory of Wireless Communication Technologies, School of Information Science and Engineering, Shandong University, Qingdao 266237, China (e-mail: changht@sdu.edu.cn).}
		\thanks{E. M. Aggoune is with the AI and Sensing Technologies Research Center, University of Tabuk, Tabuk 47315/4031, Saudi Arabia (e-mail: haggoune@ut.edu.sa).}
	}

    \markboth{IEEE Transactions on Vehicular Technology,~Vol.~XX, No.~XX, Month~2026}%
	{HOU \MakeLowercase{\textit{et al.}}: A General Non-stationary Channel Model for Autonomous Vehicle Communications in Underground Spaces}
	\maketitle

	\begin{abstract}
		\boldmath
		Autonomous vehicles exhibit high mobility and environmental adaptability, making them well-suited for underground inspection and rescue missions.		
		In this paper, a general three-dimensional (3D) non-stationary geometry based stochastic model (GBSM) is proposed for autonomous vehicle communications in underground spaces.
		The proposed channel model accounts for the 3D mobility of autonomous vehicles and can be adapted to pillar, tunnel, and cave scenarios.
		To characterize the structural and physical features of underground spaces, the proposed model incorporates the generation and evolution of spatially dispersed and/or directionally confined clusters along with their constituent specular/dense multipath components (SMCs/DMCs).	
		Additionally, the proposed channel model considers channel non-stationarity, waveguide effect, and near-field/far-field effects.
		Statistical properties, including space-time-frequency correlation functions (STF-CFs), delay/angular/Doppler power spectral densities (PSDs), root mean square (RMS) delay spread, and stationary interval, are derived and analyzed.	
		The validity of the proposed channel model is substantiated through comparative analysis between measurement and simulation results.
	\end{abstract}

\vspace{-0.3cm}
	\section{Introduction}
		\IEEEPARstart{I}{n} the pursuit of achieving global coverage for the sixth generation (6G) space-air-ground-sea integrated networks, the exploration of underground spaces has emerged as a critical and challenging frontier \cite{J23_ComST_CXWang_6G,J22_TVT_CXWang_PervasiveCM1}.	
		Underground spaces are typical harsh environments where human exploration is often perilous due to poor lighting, intricate terrain, and constricted spaces \cite{J09_ComST_SYarkan_Underground}.	 
		With high mobility and excellent adaptability, autonomous vehicles, including autonomous aerial vehicles (AAVs) and autonomous ground vehicles (AGVs), can significantly enhance operational efficiency and safety in underground detection, search, and rescue missions \cite{J21_AESM_MDePetrillo_UAVundergroundSurvey,J19_TVT_HQin_UAVundergroundSurvey}.	
		However, for autonomous vehicles operating in underground spaces, some unique characteristics arising from the complex underground spaces and the mobility of autonomous vehicles differ significantly from those in overground spaces \cite{J20_TVT_ZMa_UAVGBSMA2A, J21_IOT_HChang_UAVGBSM, J25_TWC_BHua_UAVGBSM, J25_IOT_SYang_UAVwobbling,J23_TVT_ZHuang_UAVGBSM}.	
		Therefore, it is imperative to establish an accurate and comprehensive 
		channel model to support reliable autonomous vehicle communications in underground spaces, enabling effective system design and performance evaluation.	
			
		Underground spaces generally consist of rough or tilted sidewalls and curved or uneven ceilings/floors.
		According to structural features such as length-to-width ratio and cross-sectional dimension, underground spaces can be broadly classified into pillar, tunnel, and cave scenarios \cite{J24_TCNC_BHe_UAVundergroundSurvey}.			
		Pillar scenarios feature open layouts with abundant scatterers such as supporting pillars and parked vehicles. Typical pillar scenarios include underground garages and mine pit faces.
		Tunnel scenarios have narrow and extended structures that support guided propagation, and   
		they commonly contain subway tunnels and mine roadways. 
		Cave scenarios typically exhibit irregular cross-sectional profiles and columnar deposits. Representative cave scenarios include karst caves and culverts.
		Moreover, underground spaces possess distinctive physical features such as surface roughness, wall tilt, and blockage effects. 
		Surface roughness refers to the unevenness of a surface, which is commonly described by the root mean square (RMS) value of height variations relative to an idealized reference plane~\cite{J09_TAP_WXu_Surfaceroughness}.
		Wall tilt  describes the angular deviation of a wall from the vertical plane~\cite{J75_TAP_AEmslie_Undergroundtunnel}.
		The blockage effect denotes the temporary transmission obstruction caused by humans or large machinery in underground spaces~\cite{C11_EUCAP_MJacob_Blockageeffect}.

		In recent years, extensive research on the channel measurements and analysis has been conducted in underground spaces.
		For pillar scenarios, the authors in~\cite{J19_CC_MYang_Pillarmea} conducted vehicle-to-vehicle (V2V) channel measurements in an underground garage at \mbox{5.9~GHz} to analyze the distribution of angles.
		Significant clustering of multipath components (MPCs) around~±15° in the elevation angle was observed, attributed to reflections from the floor and ceiling.
		In~\cite{J23_TAP_HChang_TunnelPillarGBSM}, channel measurements were carried out in an underground garage at 2.5~GHz and 3.5~GHz to investigate the path loss and delay spread.
		The results indicated that the path loss exponent at 3.5 GHz exceeded that at 2.5 GHz, while the delay spread exhibited minimal variation between the two frequency bands.	 
		Notably, the path loss exponents were close to 2 despite the presence of numerous obstacles because of the weak waveguide effect formed by the ceiling and floor.								
		For tunnel scenarios, multiple-input multiple-output (MIMO) channel measurements were carried out in subway tunnels at 900~MHz to investigate the effects of complicated tunnel geometric structures on the channel characteristics in~\cite{J03_JSAC_MLienard_Tunnelmea}.
		The angular spread of MPCs was found to be strongly restricted by the waveguide effect in tunnels.	
		Furthermore, railway tunnel channel measurements were conducted at 900~MHz to analyze the path loss in~\cite{J07_TVT_CBrisoRodriguez_UndergroundtunnelBP}.  The path loss model with a break point distinguishing near-field and far-field regions was proposed following the measurement data.
		In~\cite{J19_AWPL_FChallita_UndergroundDMCmea}, massive MIMO channel measurements were conducted in an underground subway	wagon at 32.5~GHz with 15 GHz bandwidth, revealing a tailing effect in the delay power spectral density (PSD).
		The specular multipath components (SMCs) and dense multipath components (DMCs) were identified with comparable power contributions, and their influence on spatial correlation was assessed.
		Additionally, V2V channel measurements in tunnel scenarios at 5.9~GHz with 30~MHz bandwidth were performed in~\cite{C20_EuCAP_SJiang_UndergroundDMCmea}, where the power of DMC  was found to occupy a small proportion under the line-of-sight (LoS) condition.
		To further investigate the influence of DMCs on system performance, massive MIMO channel measurements at 3.5~GHz with 160~MHz bandwidth were conducted in an underground subway~\cite{J22_APM_ASaleem_UndergroundDMCmea}.
		The results revealed a distance-dependent DMC power ratio due to the distinct propagation behaviors of DMCs and SMCs.	
		For cave scenarios, channel measurements at 1.3~GHz, 2.4~GHz, and 5.8~GHz in various caves were conducted in~\cite{J14_TAP_MDBedford_Cave}. Besides, theoretical models were proposed considering the additional attenuation caused by surface roughness and wall tilt. 					 			
		The above channel measurements have demonstrated that the waveguide effect and near-field/far-field effects are key propagation characteristics in underground spaces, and accurate modeling of DMCs is essential for realistic channel characterization.			
		Despite these advances, a handful of underground channel measurements consider the three-dimensional~(3D) dynamic evolution of the transmitter~(Tx) and receiver~(Rx).

		Generally, existing channel models can be classified into deterministic and stochastic categories.
		Given the unique channel characteristics of underground spaces, deterministic models primarily include those based on waveguide theory, vector parabolic equation (VPE), and ray tracing (RT).
		In~\cite{J10_TCOM_ZSun_Undergroundtunnelpillar}, an extended multimode waveguide model was proposed to characterize wave propagation within both tunnel and pillar scenarios, and the effects of different frequencies, space dimensions, antenna parameters on the signal propagation were analyzed.	
		In~\cite{J19_TAP_XZhang_Undergroundtunnel}, a VPE-based model incorporating surface roughness was proposed for tunnel scenarios, and the path loss were analyzed to evaluate the impact of surface roughness.
		In~\cite{C21_WCNC_MWu_Undergroundpillarroughness}, a RT channel model for pillar scenarios was proposed to investigate how surface roughness influences signal propagation, with particular attention given to analyzing the RMS delay spread.
		While deterministic models can characterize propagation in underground spaces, they suffer from high computational complexity.
				
		Among stochastic channel models, geometry based stochastic models (GBSMs) have gained widespread adoption owing to their flexibility and scalability in representing complex multipath propagation, as well as their good agreement with measurement data~\cite{J25_SCIS_CXWang_Enhanced6G}.
		For instance, a semi-circular GBSM tailored for V2V communications in tunnel scenarios was proposed in~\cite{J16_TVT_NAvazov_TunnelGBSM}, assuming single-bounce propagation from scatterers randomly distributed along tunnel walls.			
		In~\cite{J20_TITS_YLiu_TunnelGBSM}, a 3D non-stationary GBSM for high-speed train communications in cuboid tunnel scenarios was proposed.
		Based on the geometrical relationship, the positions of clusters with single-bounce and multi-bounce components were determined by time-varying angular information.	
        In~\cite{C23_ICCC_XJi_SurfaceroughnessBlockageGBSM}, a GBSM for underground mines communications was proposed by incorporating surface roughness and human blockage, and their effects on spatial correlation and channel capacity were analyzed.	 
		In~\cite{J23_TAP_HChang_TunnelPillarGBSM}, a GBSM for vehicular communications applicable to both tunnel and pillar scenarios was proposed. 
		By leveraging the twin cluster model framework, the generation and evolution at the cluster level were modeled to describe the channel characteristics. 				
		However, the aforementioned channel models are mainly limited to tunnel and pillar scenarios, neglecting cave scenarios and lacking a general channel model for diverse underground scenarios. 
		Moreover, they neither characterize DMCs nor support 3D mobility of transceiver in underground spaces. 	
		While several studies have extended GBSMs to incorporate DMCs for indoor office environment\mbox{\cite{C15_EuCAP_UTVirk_DMCgbsm,C21_PIMRC_YLi_DMCgbsm}} and industrial Internet of things scenarios \cite{J25_TVT_LZhang_DMCgbsm}, channel modeling of DMCs for autonomous vehicle communications in underground spaces remains unexplored. 			
		
		To address the above limitations, we propose a general 3D non-stationary channel model for autonomous vehicle communications in underground spaces.
		This model accounts for both the 3D mobility of autonomous vehicles and the unique features of underground spaces. 
		By leveraging the structural and physical features of underground spaces, the model integrates the generation and evolution mechanisms of clusters and rays.
		The main contributions and novelties of this paper are summarized as follows.
		\begin{itemize}
			\item 
			A general 3D non-stationary MIMO GBSM is proposed for autonomous vehicle communications in underground spaces.
			A posture variation matrix is employed to characterize the 3D mobility of autonomous vehicles.
			The channel non-stationarity induced by 3D mobility, along with key propagation characteristics such as the waveguide effect and near-field/far-field effects, is modeled to reflect realistic propagation in underground spaces.													
			\item 		
			The generation and evolution mechanisms of clusters and rays are formulated to accurately depict scattering environment.
			Based on structural features, spatially dispersed and/or directionally confined clusters are generated and updated.	
			Based on physical features, DMCs and SMCs are modeled with power decay described by a cosine-squared function that incorporates a roughness attenuation factor.			
			By adjusting cluster parameters and the DMC power ratio, the proposed channel model is applicable to pillar, tunnel, and cave scenarios.
			\item  
			Statistical properties, including space-time-frequency correlation functions (STF-CFs), delay/angular/Doppler PSDs, RMS delay spread, and stationary interval, are analytically derived and numerically simulated.
			Moreover, comparison between measured and simulated results in different scenarios are analyzed.
		\end{itemize}	
	
		The rest of this paper is organized as follows. 
		Section~\ref{p2} presents the proposed channel model for autonomous vehicle communications in underground spaces.
		Section~\ref{p3} derives the statistical properties of the model. Section~\ref{p4} provides an analysis of simulation results in different scenarios. 
		Finally, Section~\ref{p5} concludes the paper.
		
	\section{Autonomous Vehicles Channel Model for Underground Spaces}\label{p2}
\begin{figure}[!t]		
			\centering	{\includegraphics[width=3.3in]{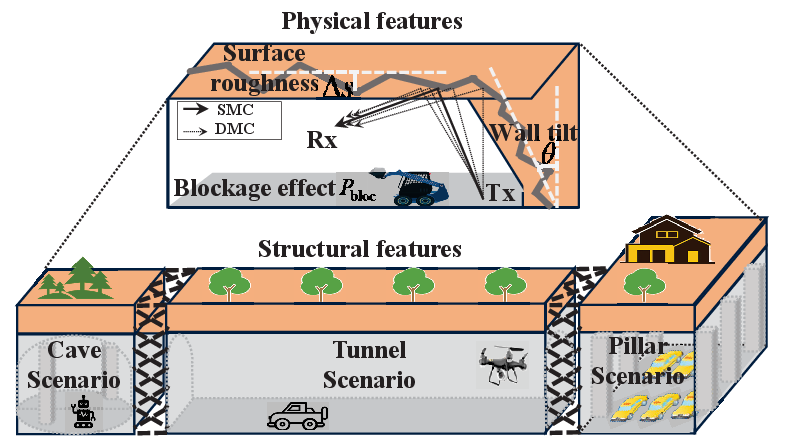}}		
			\captionsetup{justification=raggedright, singlelinecheck=false}
			\caption{Structural and physical features of underground spaces.}
			\label{UG}
		\end{figure}

		The structural and physical features of underground spaces are shown in Fig.~\ref{UG}.
		Underground spaces can be abstracted as the cuboid-shaped structure with dimensions of length~$l$, width~$w$, and height~$h$.		
		Based on the length-to-width ratio, i.e., $l/w$, and cross-sectional dimension, i.e., $w \times h$, underground spaces can be classified into tunnel, cave, and pillar scenarios.
		The physical features of underground spaces include surface roughness represented by~$\Delta s$, wall tilt denoted by~$\theta$, and the blockage effects quantified by the blockage probability ~${P_{{\rm{bloc}}}}$.
		Due to the confined spaces, electromagnetic waves inside underground spaces experience reflections and diffuse scattering, resulting in severe fading. As depicted in Fig.~\ref{UG}, strong MPCs generated by specular reflections from large and smooth surfaces constitute SMCs, while numerous weak diffuse scattering MPCs induced by surface roughness and wall tilt superpose to form DMCs~\cite{J10_AWPL_JPoutanen_DMCmedel}.

		\begin{figure*}[!t]			
			\centering	{\includegraphics[width=6.6in]{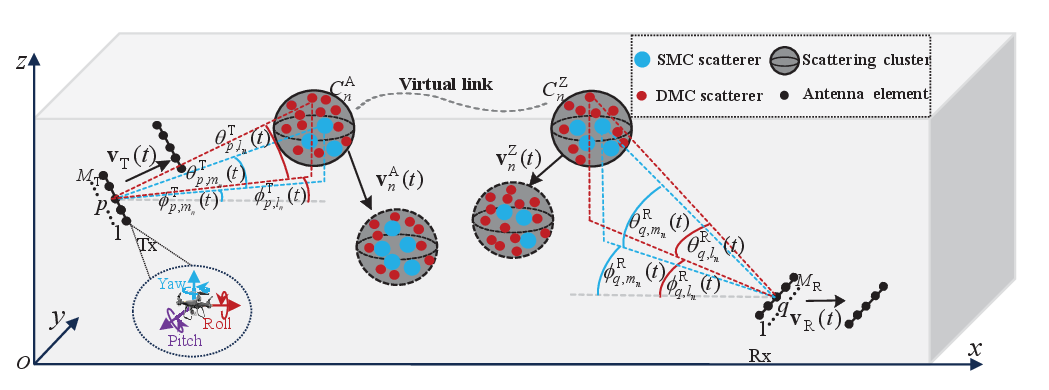}}
			\captionsetup{justification=raggedright, singlelinecheck=false}
			\caption{Illustration of autonomous vehicles channel model in underground spaces.}
			\label{CM}
		\end{figure*}
				\begin{figure*}[!t]		
			\centering	{\includegraphics[width=6.6in]{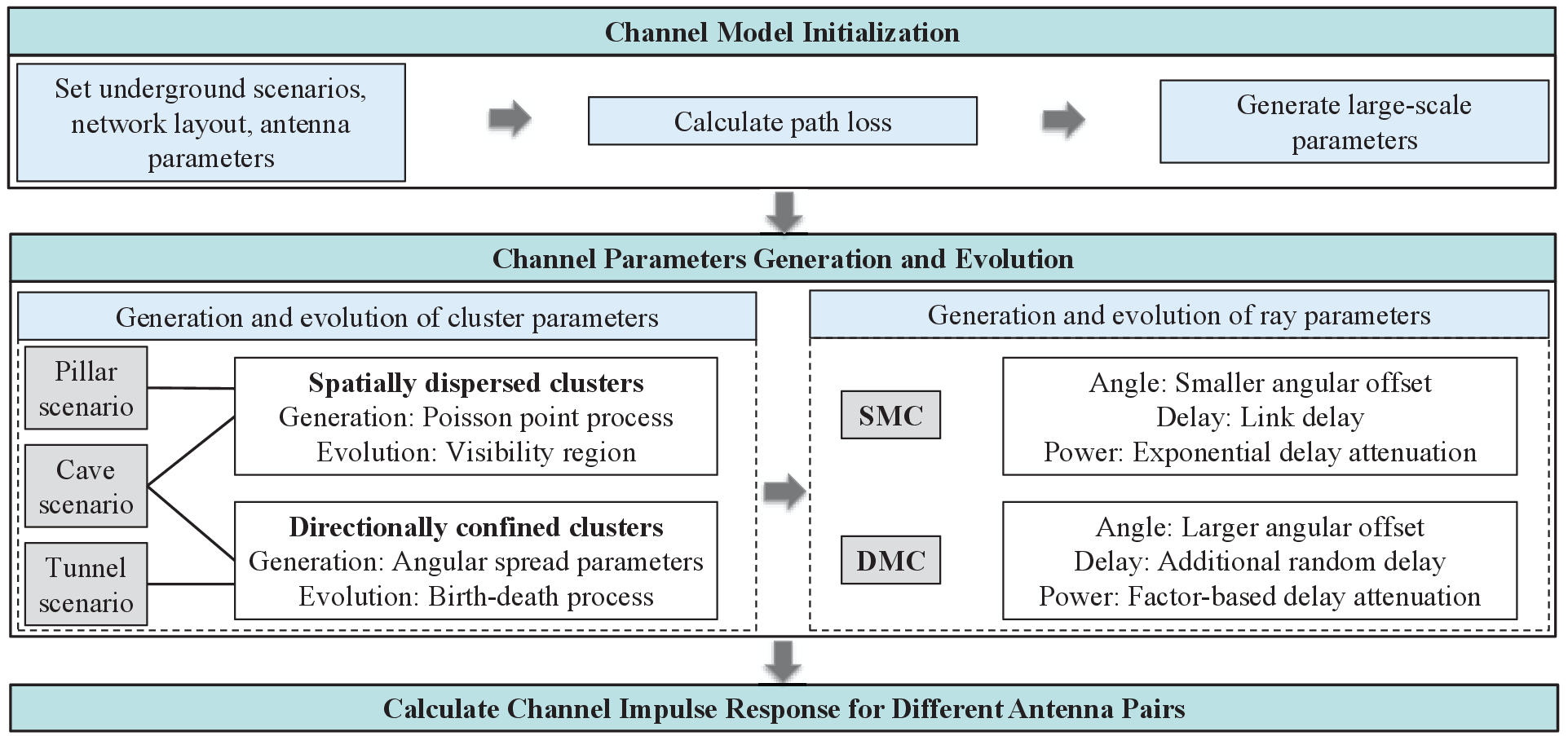}}		
			\captionsetup{justification=raggedright, singlelinecheck=false}
			\caption{Channel coefficients generation process.}
			\label{CC}
		\end{figure*}
		Taking the distinctive 3D mobility of autonomous vehicles and physical features of underground spaces into account, a general non-stationary MIMO autonomous vehicles GBSM in underground spaces is constructed. As shown in Fig.~\ref{CM}, an AAV equipped with $M_{\rm{T}}$ antenna elements acts as the Tx, and an AGV equipped with $M_{\rm{R}}$ antenna elements serves as the Rx.
		At time instant $t$, the AAV moves in 3D underground spaces with the velocity vector ${{\bf{v}}_{\rm{T}}}(t)=v_{\rm{T}}(t)[{\rm cos}\xi_{\rm{T}}{\rm cos}\gamma_{\rm{T}}, {\rm sin}\xi_{\rm{T}}{\rm cos}\gamma_{\rm{T}},{\rm sin}\gamma_{\rm{T}}]$, and AGV moves in two-dimensional (2D) underground floor with the velocity vector \mbox{${{\bf{v}}_{\rm{R}}}(t) = v_{\rm{R}}(t)[{\rm cos}\xi_{\rm{R}}, {\rm sin}\xi_{\rm{R}},0]$}, where $\xi_{\rm{T}}$ and $\gamma_{\rm{T}}$ stand for the azimuth and elevation angles of velocity of Tx, and  $\xi_{\rm{R}}$ is the azimuth angle of velocity of Rx.
		The \mbox{$p$-th ($p=1, \cdots, M_{\rm{T}}$)} Tx antenna element is  denoted by $A_p^{\rm{T}}$, and the \mbox{$q$-th ($q=1, \cdots, M_{\rm{R}}$)} Rx antenna element is  denoted by $A_q^{\rm{R}}$. 
		Considering the attitude variations of AAV,  the vector of $A_p^{\rm{T}}$ in the global coordinate system (GCS) is expressed as ${\bf{A}}_p^{{\rm{T}}}(t) = {{\bf{R}}}(t){\bf{A}}_p^{\rm{'T}}(t)$. Here, ${\bf{A}}_p^{\rm{'T}}(t)$ is the vector of $A_p^{\rm{T}}$ in the local coordinate system (LCS) and ${{\bf{R}}}(t)$ is the posture variation matrix represented as \cite{3GPP38.901}
		\begin{equation}\label{Rotation}
			\begin{array}{l}
				{{\bf{R}}}(t) = \left[\hspace{-0.2cm} {\begin{array}{*{20}{c}}
						{\cos {\varphi }(t)}&{ - \sin {\varphi }(t)}&0\\
						{\sin {\varphi }(t)}&{\cos {\varphi}(t)}&0\\
						0&0&1
				\end{array}} \hspace{-0.2cm}\right]\\
				\cdot \left[\hspace{-0.2cm} {\begin{array}{*{20}{c}}
						{\cos {\gamma }(t)}&0&{\sin {\gamma}(t)}\\
						0&1&0\\
						{ - \sin {\gamma}(t)}&0&{\cos {\gamma }(t)}
				\end{array}} \hspace{-0.2cm}\right]\left[\hspace{-0.2cm} {\begin{array}{*{20}{c}}
						1&0&0\\
						0&{\cos {\beta }(t)}&{ - \sin {\beta}(t)}\\
						0&{\sin {\beta }(t)}&{\cos {\beta}(t)}
				\end{array}} \hspace{-0.2cm}\right]
			\end{array}
		\end{equation}where $\gamma(t)$, $\varphi(t)$, and $\beta(t)$ represent the roll, yaw, and pitch angles defined in the LCS with respect to the GCS.
		The channel model can switch between AAV and AGV mobility by assigning a 3D velocity vector and posture matrix or a floor-constrained 2D velocity vector to each endpoint, respectively.
		To flexibly represent multi-bounce propagation in enclosed underground spaces, scatterers originating from objects and walls are conceptualized as twin clusters. 
		Here, the clusters correspond to directionally confined clusters and/or spatially dispersed clusters, which are characterized by structural features.
		Each twin cluster comprises a first-bounce cluster $C_{n}^{\rm{A}}$ near the Tx, a last-bounce cluster $C_{n}^{\rm{Z}}$ near the Rx, and a virtual link representing the intervening propagation space. 
		It is noteworthy that when the delay associated with the virtual link is set to zero, the twin cluster degenerates into a single cluster, thereby facilitating the modeling of single-bounce propagation.
		At time instant $t$, the communication channel comprises a total of $N(t)$ clusters.	
		Each cluster contains scatterers responsible for generating SMC and DMC rays.
		To accurately  depict the propagation environment, scatterers generating DMCs are assumed smaller and more numerous than scatterers producing SMCs. 
		Within the $n$-th cluster, there exist ${M_n}(t)$ SMC rays and ${L_n}(t)$ DMC rays, respectively.		
		Both the rays of SMC and DMC share the same cluster center but exhibit distinct multipath parameters.		
		For illustrative purposes,  Fig.~\ref{CM} depicts solely the \mbox{$n$-th} twin cluster.
		Key parameters associated with the channel model, along with their definitions, are summarized in Table~\ref{Tabparameters}.
		Furthermore, Fig.~\ref{CC} provides an overview of the channel coefficient generation process.
		\begin{center}	
			\begin{table*}[h]	
				\caption{Definitions of Parameters in the Proposed Channel Model. }
				\begin{center}
					\begin{spacing}{1.19}
						\begin{tabular}{|c|c|}
							\hline
							\textbf{Parameters} & \textbf{Definitions}\\
							\hline
							$M_{\rm{T(R)}}$, $\delta_{\rm{T(R)}}$ &  The element number and spacing of Tx(Rx) antenna array\\
							\hline
							$\alpha_{\rm{T(R)}}$, $\beta_{\rm{T(R)}}$ &  The azimuth and elevation angles of Tx(Rx) antenna array\\						
							\hline
							$A_p^{\rm{T}}$, $A_q^{\rm{R}}$ &  The $p$-th Tx antenna element and $q$-th Rx antenna element\\
							\hline
							$C_{n}^{\rm{A}}$, $C_{n}^{\rm{Z}}$ & The first and last bounce for the $n$-th cluster\\
							\hline
							${{\bf{v}}_{\rm{T}}}(t)$, ${{\bf{v}}_{\rm{R}}}(t)$, ${\bf{v}}_{n}^{\rm{A}}(t)$, ${\bf{v}}_{n}^{\rm{Z}}(t)$ &  The velocity vectors of Tx, Rx, $C_{n}^{\rm{A}}$, and $C_{n}^{\rm{Z}}$\\							
							\hline
							$N(t)$ &  The time-variant number of clusters\\
							\hline
							${M_n}(t)$, ${L_n}(t)$ &  The number of SMC and DMC rays in the $n$-th cluster\\					
							\hline
							$\phi _{p(q)}^{{\rm{LoS}}}(t)$ &  The azimuth angle of departure/arrival (AAoD/AAoA) of LoS path for $A_p^{\rm{T}}$($A_q^{\rm{R}}$)\\
							\hline
							$\theta _{p(q)}^{{\rm{LoS}}}(t)$ &  The elevation angle of  departure/arrival (EAoD/EAoA) of LoS path for $A_p^{\rm{T}}$($A_q^{\rm{R}}$)\\
							\hline
							$\phi _{{q}{\rm{,}}{m_n}}^{\rm{R}}(t)$, $\theta _{{q}{\rm{,}}{m_n}}^{\rm{R}}(t)$ &   The AAoA and EAoA of $m$-th SMC ray in $C_{n}^{\rm{Z}}$ for $A_q^{\rm{R}}$\\
							\hline
							$\phi _{{p}{\rm{,}}{m_n}}^{\rm{T}}(t)$, $\theta _{{p}{\rm{,}}{m_n}}^{\rm{T}}(t)$ &   The AAoD and EAoD of $m$-th SMC ray in $C_{n}^{\rm{A}}$ for $A_p^{\rm{T}}$\\
							\hline
							$\phi _{{q}{\rm{,}}{l_n}}^{\rm{R}}(t)$, $\theta _{{q}{\rm{,}}{l_n}}^{\rm{R}}(t)$ &  The AAoA and EAoA  of $l$-th DMC ray in $C_{n}^{\rm{Z}}$ for $A_q^{\rm{R}}$\\
							\hline	
							$\phi _{{p}{\rm{,}}{l_n}}^{\rm{T}}(t)$, $\theta _{{p}{\rm{,}}{l_n}}^{\rm{T}}(t)$ &  The AAoD and EAoD of $l$-th DMC ray in $C_{n}^{\rm{A}}$ for $A_p^{\rm{T}}$\\
							\hline			
						\end{tabular}
					\end{spacing}			
					\vspace{-\baselineskip}
					\label{Tabparameters}				
				\end{center}
			\end{table*}
		\end{center}
	\vspace{-0.5cm}
		\subsection{Channel Impulse Response}
				Considering the complex propagation characteristics of electromagnetic waves in underground spaces, the channel impulse response (CIR) matrix for the autonomous vehicles channel  can be expressed as \cite{J22_TVT_CXWang_PervasiveCM1}
				\begin{equation}\label{H}
						{\bf{H}} ={[PL \cdot SH]^{1/2}} \cdot {{\bf{H}}_{\rm{S}}}
				\end{equation}where $PL$ represents the path loss, $SH$ denotes the shadowing, and ${{{{\bf{H}}_{\rm{S}}}} }$ defines the matrix of small-scale fading.
				To account for the transition from near-field region dominated by free-space propagation to far-field region dominated by multimode waveguide propagation in underground space, we adopt a two-segment path loss model 
				\begin{equation}\label{PL}
					PL = \left\{ {\begin{array}{*{20}{c}}{10\alpha_1{{\log}_{10}}\left( d \right) + \beta_1 }&{,d \le {d_{b}}}\\[3pt]
					{10\alpha_2{{\log }_{10}}\left( {{d_{b}}} \right) +\beta_2 +\gamma \left( {d - {d_{b}}} \right)}&{,d > {d_{b}}}
					\end{array}} \right.
				\end{equation}
				where $d$ is the distance between Tx and Rx, ${d_{b}}$ is the boundary between the near-field and far-field regions.
				This boundary depends on the cross-sectional dimensions of underground spaces as well as the relative positions of the Tx and Rx~\cite{C11_VTC_KGuan_UndergroundtunnelBP}.
				Furthermore, $\beta_1$ and $\beta_2$ correspond to the intercepts, while $\alpha_1$ and $\alpha_2$ denote the path loss exponents within the near-field and far-field regions, respectively.
				Parameters $\alpha_1$, $\alpha_2$, $\beta_1$, and $\beta_2$ in different underground scenarios can be obtained from experimental channel measurements~\cite{J23_TAP_HChang_TunnelPillarGBSM,J07_TVT_CBrisoRodriguez_UndergroundtunnelBP,J07_TAP_MRak_TunnelCave}.		
				Moreover, the additional attenuation rate in dB/m caused by lossy waveguide, surface roughness and wall tilt can be represented as \cite{J14_TAP_MDBedford_Cave}
				\begin{equation}\label{gam}
					\gamma  = {\gamma _{{\rm{mode}}}} + {\gamma _{{\rm{roug}}}} + {\gamma _{{\rm{tilt}}}}
				\end{equation}
			  	where ${\gamma _{{\rm{mode}}}} = \frac{{{\lambda ^2}}}{2}\left( {\frac{1}{{{w^3}\sqrt {{\varepsilon _R} - 1} }} + \frac{{{\varepsilon _R}}}{{{h^3}\sqrt {{\varepsilon _R} - 1} }}} \right)$, $	{\gamma _{{\rm{roug}}}} = \frac{{{\pi ^2}{\Delta s}^2\lambda }}{2}\left( {\frac{1}{{{w^4}}} + \frac{1}{{{h^4}}}} \right)$, and ${\gamma _{{\rm{tilt}}}} = \frac{{{\pi ^2}{\theta ^2}}}{{2\lambda }}$. 
				Here, $\lambda$ is the wavelength,  ${\varepsilon_R}$ is the relative permittivity. 
				It is noteworthy that the boundary  ${d_{b}}$ not only distinguishes regions exhibiting different large-scale fading behaviors but also corresponds to variations in small-scale fading characteristics.

			    In underground spaces, small-scale fading of autonomous vehicles channel arises primarily due to complex multipath propagation.
			    The small-scale fading CIR between $A_q^{\rm{R}}$ and $A_p^{\rm{T}}$ is given by ${h_{qp}}(t,\tau )={\left[ {{{\bf{H}}_{\rm{S}}}} \right]_{qp}}$, and can be written~as  
			    \begin{equation}\label{h}
				   \begin{array}{l}
						\!{h_{qp}}(t,\tau ) = \sqrt {\frac{{{K_R}(t)}}{{{K_R}(t) + 1}}}h_{qp}^{{\rm{LoS}}}(t,\tau )\\[5pt]
						\!+\! \sqrt {\frac{{1}}{\!{{K_R}(t) + 1}}}\! \left[\! {\!\sqrt{{\mu}(t)} h_{qp}^{{\rm{DMC}}}\!(t,\tau )\! +\! \sqrt {1 \!-\! {\mu}(t)} h_{qp}^{{\rm{SMC}}}\!(t,\tau )} \!\right]\\
					\end{array}
			    \end{equation}where ${K_R}(t)$ is the Ricean K-factor, ${\mu}(t) = \frac{{P_{{\rm{DMC}}}(t)}}{{P_{{\rm{SMC}}}(t){\rm{ + }}P_{{\rm{DMC}}}(t)}}$ denotes the DMC power ratio.	    
			   	Different underground scenarios can be characterized by adjusting cluster parameters and the DMC power ratio. This flexibility enables the model to accurately reflect the channel non-stationarity and the propagation characteristics of underground spaces. 	
			   		   	
			   	The channel coefficient of LoS component is given as
			    \begin{equation}\label{los}
			    	\begin{array}{l}
						h_{qp}^{{\rm{LoS}}}\!(t,\tau ){\kern 1pt}  \!=\! {\left[ {\begin{array}{*{20}{c}}
									{\hspace{-0.2cm}{F_{q,{\rm{V}}}}(\theta _q^{{\rm{LoS}}}(t),\phi _q^{{\rm{LoS}}}(t))\hspace{-0.2cm}}\\
									{\hspace{-0.2cm}{F_{q,{\rm{H}}}}(\theta _q^{{\rm{LoS}}}(t),\phi _q^{{\rm{LoS}}}(t))\hspace{-0.2cm}}
							\end{array}} \right]^{\rm{T}}}\left[ {\begin{array}{*{20}{c}}
								{\hspace{-0.2cm}{e^{{\rm{j}}\Theta _{{\rm{LoS}}}^{{\rm{VV}}}}}\hspace{-0.2cm}}&0\\
								0&{\hspace{-0.2cm}{e^{{\rm{j}}\Theta _{{\rm{LoS}}}^{{\rm{HH}}}}}\hspace{-0.2cm}}
						\end{array}} \right]\\[10pt]
					\!\cdot \!\left[ {\begin{array}{*{20}{c}}
								{\hspace{-0.2cm}{F_{p,{\rm{V}}}}(\theta _p^{{\rm{LoS}}}(t),\phi _p^{{\rm{LoS}}}(t))\hspace{-0.2cm}}\\
								{\hspace{-0.2cm}{F_{p,{\rm{H}}}}(\theta _p^{{\rm{LoS}}}(t),\phi _p^{{\rm{LoS}}}(t))\hspace{-0.2cm}}
						\end{array}} \right]\!{e^{{\rm{j}}2\pi {f_{\rm{c}}}\tau _{qp}^{{\rm{LoS}}}(t)}} \cdot \!\delta \!\left( {\tau \! -\! \tau _{qp}^{{\rm{LoS}}}(t)} \!\right)
					\end{array}
			    \end{equation}where $f_{\rm{c}}$ is the carrier frequency, $\tau _{qp}^{{\rm{LoS}}}(t) = {{\left\| {{\bf{A}}_p^{\rm{T}}(t) - {\bf{A}}_q^{\rm{R}}(t)} \right\|} \mathord{\left/
			    {\vphantom {{\left\| {{\bf{A}}_p^{\rm{T}}(t) -{\bf{A}}_q^{\rm{R}}(t)} \right\|} c}} \right.\kern-\nulldelimiterspace} c}$ denotes the delay of LoS component with $c$ representing the speed of light and ${\bf{A}}_q^{\rm{R}}(t)$ representing the vector of $A_q^{\rm{R}}$ in the GCS.
		    	The random phases $\Theta_{\text{LoS}}^{\rm{VV}}$ and $\Theta_{\text{LoS}}^{\rm{HH}}$ are uniformly distributed over (0, $2\pi$].
		        Symbols $F_{p(q),{\rm{V}}}$ and $F_{p(q),{\rm{H}}}$ denote the vertical and horizontal polarization antenna patterns corresponding to $A_p^{\rm{T}}$ ($A_q^{\rm{R}}$), respectively.			    	
			    		    		
			    The channel coefficients of SMCs and DMCs are given as	     
				\begin{equation}\label{M_SMC}
					\begin{array}{l}
						h_{qp}^{{\rm{SMC}}}(t,\tau ) = \sum\limits_{n = 1}^{N(t)}{\sum\limits_{{m_n} = 1}^{{M_n}(t)} {{{\left[\hspace{-0.2cm} {\begin{array}{*{20}{l}}
												{{F_{q,{\rm{V}}}}\left(\! {\theta_{q{\rm{,}}{m_n}}^{\rm{R}}(t),\phi _{q{\rm{,}}{m_n}}^{\rm{R}}(t)} \!\right)}\\
												{{F_{q,{\rm{H}}}}\left(\! {\theta_{q{\rm{,}}{m_n}}^{\rm{R}}(t),\phi _{q{\rm{,}}{m_n}}^{\rm{R}}(t)}\! \right)}
										\end{array}}\hspace{-0.2cm}\right]}^{\rm{T}}}} } \\[15pt]
								\!	\cdot	\!\left[\! \hspace{-0.2cm}{\begin{array}{*{20}{c}}
								\!{\!{e^{{\rm{j}}\Theta_{\!{m_n}}^{\!{\rm{VV}}}}}}&\!{\!\sqrt {\kappa _{\!{m_n}}^{\! - 1}} {e^{{\rm{j}}\Theta _{\!{m_n}}^{\!{\rm{VH}}}}}}\\
								{\sqrt {\kappa _{\!{m_n}}^{\! - 1}}{e^{{\rm{j}}\Theta _{\!{m_n}}^{\!{\rm{HV}}}}}}&{\!{e^{{\rm{j}}\Theta _{\!{m_n}}^{\!{\rm{HH}}}}}}
						\end{array}}\! \hspace{-0.2cm}\right]\!\left[\!\hspace{-0.2cm}{\begin{array}{*{20}{l}}
								{{F_{p,{\rm{V}}}}\!\left(\! {\theta_{p{\rm{,}}{m_n}}^{\rm{T}}\!(t),\!\phi _{p{\rm{,}}{m_n}}^{\rm{T}}\!(t)} \!\right)}\\
								{{F_{p,{\rm{H}}}}\!\left(\! {\theta_{p{\rm{,}}{m_n}}^{\rm{T}}\!(t),\!\phi _{p{\rm{,}}{m_n}}^{\rm{T}}\!(t)} \!\right)}
						\end{array}} \hspace{-0.2cm}\!\right]\\[15pt]
					\cdot	\sqrt {P_{qp,{m_n}}(t)}  \cdot {e^{{\rm{j}}2\pi{f_{\rm{c}}}\tau _{qp,{m_n}}(t)}} \cdot \delta \left( {\tau  - \tau _{qp,{m_n}}(t)} \right)
					\end{array}
				\end{equation}
				\begin{equation}\label{M_DMC}
					\begin{array}{l}
						h_{qp}^{{\rm{DMC}}}(t,\tau ) = \sum\limits_{n = 1}^{N(t)}{\sum\limits_{{l_n} = 1}^{{L_n}(t)} {{{\left[\hspace{-0.2cm} {\begin{array}{*{20}{l}}
												{{F_{q,{\rm{V}}}}\left(\! {\theta_{q{\rm{,}}{l_n}}^{\rm{R}}(t),\phi _{q{\rm{,}}{l_n}}^{\rm{R}}(t)} \!\right)}\\
												{{F_{q,{\rm{H}}}}\left(\! {\theta_{q{\rm{,}}{l_n}}^{\rm{R}}(t),\phi _{q{\rm{,}}{l_n}}^{\rm{R}}(t)}\! \right)}
										\end{array}}\hspace{-0.2cm}\right]}^{\rm{T}}}} } \\[15pt]
							\cdot	\!\left[\! \hspace{-0.2cm}{\begin{array}{*{20}{c}}
								{{e^{{\rm{j}}\Theta_{\!{l_n}}^{\!{\rm{VV}}}}}}&\!{\sqrt {\kappa _{\!{l_n}}^{\! - 1}} {e^{{\rm{j}}\Theta _{\!{l_n}}^{\!{\rm{VH}}}}}}\\
								{\sqrt {\kappa _{\!{l_n}}^{\! - 1}}{e^{{\rm{j}}\Theta _{\!{l_n}}^{\!{\rm{HV}}}}}}&{\!{e^{{\rm{j}}\Theta _{\!{l_n}}^{\!{\rm{HH}}}}}}
						\end{array}}\! \hspace{-0.2cm}\right]\!\left[\!\hspace{-0.2cm}{\begin{array}{*{20}{l}}
								{{F_{p,{\rm{V}}}}\!\left(\! {\theta_{p{\rm{,}}{l_n}}^{\rm{T}}\!(t),\!\phi _{p{\rm{,}}{l_n}}^{\rm{T}}\!(t)} \!\right)}\\
								{{F_{p,{\rm{H}}}}\!\left(\! {\theta_{p{\rm{,}}{l_n}}^{\rm{T}}\!(t),\!\phi _{p{\rm{,}}{l_n}}^{\rm{T}}\!(t)} \!\right)}
						\end{array}} \hspace{-0.2cm}\!\right]\\[15pt]
						\cdot	\sqrt {P_{qp,{l_n}}(t)}  \cdot {e^{{\rm{j}}2\pi{f_{\rm{c}}}\tau _{qp,{l_n}}(t)}} \cdot \delta \left( {\tau  - \tau _{qp,{l_n}}(t)} \right)
					\end{array}
			    \end{equation}where $\Theta _{{m_n}({l_n})}^{{\rm{VV}}}$, $\Theta _{{m_n}({l_n})}^{{\rm{VH}}}$, $\Theta _{{m_n}({l_n})}^{{\rm{HV}}}$, and $\Theta _{{m_n}({l_n})}^{{\rm{HH}}}$ are random phases  uniformly distributed over (0, $2\pi$], $\kappa _{{m_n}({l_n})}$ is the cross polarization power ratio \cite{J22_APM_ASaleem_UndergroundDMCmea}.
		 	    The time-varying delays and powers are denoted by $\tau _{qp,{m_n}}(t)$ and $P_{qp,{m_n}}(t)$ for SMCs, and $\tau _{qp,{l_n}}(t)$ and $P_{qp,{l_n}}(t)$ for DMCs, which will be elaborated in Section \ref{DMCSMC}.	 
		 	    Notably, DMC parameters are defined in the same way as the SMC parameters, but with distinct  distributions and values.

		 	   \vspace{-0.3cm} 
		\subsection{Generation and Evolution of Cluster Parameters} 
			 Structural features significantly shape the spatial distribution of clusters in underground spaces. 
			 For the pillar scenario, abundant scatterers and a spacious layout within the finite space lead to spatially dispersed clusters.	
			 For the tunnel scenario, the narrow and extended structure concentrates clusters along the tunnel axis, resulting in directionally confined clusters.
			 For the cave scenario, characterized by structural features intermediate between tunnel and pillar scenarios, both directionally confined clusters and spatially dispersed clusters coexist.			 
			 Therefore, to accurately characterize the channel non-stationarity induced by autonomous vehicle mobility and the propagation characteristics of underground spaces such as the waveguide effect and near-field/far-field effects, the proposed channel model integrates the generation and evolution mechanisms of clusters.
			 
		 
		 	\subsubsection{Pillar Scenario}
			 	In the pillar scenario, spatially distributed clusters are modeled using a Poisson point process (PPP). 
			 	Specifically, consider a finite pillar scenario characterized by a volume \mbox{$V = w \times h \times l$} and a scattering cluster density denoted as
			 	${\lambda_{\rm{P}}}$.
			 	Under such conditions, the total number of scattering clusters follows a Poisson distribution with parameter ${\lambda_{\rm{P}}}$, i.e.,  $I_{{\rm{total}}}={\rm{Poisson}}({\lambda _{\rm{P}}}{V})$~\cite{J23_TAP_HChang_TunnelPillarGBSM}. 
			 	The 3D coordinate set of clusters center is represented as ${\bf{C}^{\rm{P}}} = [{\bf{C}}_1^{\rm{P}}, \cdots ,{\bf{C}}_i^{\rm{P}}, \cdots ,{\bf{C}}_{{I_{{\rm{total}}}}}^{\rm{P}}]$.
			 	Given that scatterers located in the far zone may be occluded by those in the near zone, scatterers in the vicinity of the Tx and Rx  should be incorporated into channel simulations.
			 	To this end, a visibility region (VR) with radius $R_{\rm  lim}$ is defined to identify effective clusters.
			 	Clusters situated within the VR of the Tx and Rx are deemed effective for multipath formation.
			 	At time instant $t$, the clusters satisfying $\left\| {{{\bf{C}}_i^{\rm{P}}} - {\bf{A}}_p^{{\rm{T}}}(t)} \right\| \le R_{\rm  lim}$ are determined as effective first-bounce cluster set ${{\bf{C}}^{\rm{A}}(t)}=[{\bf{C}}_1^{\rm{A}}(t), \cdots ,{\bf{C}}_n^{\rm{A}}(t), \cdots ,{\bf{C}}_{N(t)}^{\rm{A}}(t)]$, and the cluster satisfying  $\left\| {{{\bf{C}}_i^{\rm{P}}} - {\bf{A}}_q^{{\rm{R}}}(t)} \right\| \le R_{\rm  lim}(t)$ are determined as effective last-bounce cluster set ${{\bf{C}}^{\rm{Z}}(t)}=[{\bf{C}}_1^{\rm{Z}}(t), \cdots ,{\bf{C}}_n^{\rm{Z}}(t), \cdots ,{\bf{C}}_{N(t)}^{\rm{Z}}(t)]$.
			 	Subsequently, sub-clusters that are successfully matched between these sets merge to form effective clusters, which contribute to the channel response.
			 				 
			 	A time-variant visibility analysis is performed for all existing clusters, determining their activity status with respect to the Tx and Rx.
			 	With the movement of transceiver, clusters outside the VR are deactivated, while new emerging clusters emerging within the VR are activated.
			 	This dynamic process facilitates an accurate depiction of cluster evolution over time. 

			\subsubsection{Tunnel Scenario}
				In the tunnel scenario, the mean number of directionally confined clusters is controlled by the generation rate $\lambda_{\rm{G}}$ and recombination rate $\lambda_{\rm{R}}$ as
				\begin{equation}
					\mathbb{E}\left[ {N(t)} \right] =\frac{{{\lambda_{\rm{G}}}}}{{{\lambda _{\rm{R}}}}}
			    \end{equation}where $\mathbb{E}\left[ \cdot \right]$ is the expectation.
		        Considering near-field/far-field effects, the scattering clusters are generated with different generation and recombination rates and angular spreads in near-field and far-field regions.     
		        Specially, the AAoD and EAoD from $A_p^{\rm{T}}$ to the center of $C_{n}^{\rm{A}}$ are given as 	
				\begin{equation}
				 	{\phi}_{p,n}^{\rm{T}}={\rm AS}\left[{\phi}_{p,n}^{\rm{T}}\right]Y_{p,n}^{\rm{A}}+{\Psi}_{p,n}^{\rm{A}}
				\end{equation}	
			 	\begin{equation}
			 		{\theta}_{p,n}^{\rm{T}}={\rm AS}\left[{\theta}_{p,n}^{\rm{T}}\right]Y_{p,n}^{\rm{E}}+{\Psi}_{p,n}^{\rm{E}}
			 	\end{equation}		
				where ${\rm AS}\left[{\phi}_{p,n}^{\rm{T}}\right]$ and ${\rm AS}\left[{\theta}_{p,n}^{\rm{T}}\right]$ denote the angle spreads, ${\Psi}_{p,n}^{\rm{A}}$ and ${\Psi}_{p,n}^{\rm{E}}$ are the mean values of AAoD and EAoD, and $Y_{p,n}^{\rm{A}}$ and $Y_{p,n}^{\rm{E}}$ follow the normal distribution $N(0,1)$.
				Similarly, the  AAoA and EAoA from $A_q^{\rm{R}}$ to the center of $C_{n}^{\rm{Z}}$, i.e., ${\phi}_{q,n}^{\rm{R}}$ and ${\theta}_{q,n}^{\rm{R}}$, can be obtained in the same way.				
				Owing to the spatial limitations within the tunnel scenario, it is necessary to calculate the distance from the scattering cluster to the transceiver according to the width and height of tunnel.
				The angles of the projected paths from $A_p^{\rm{T}}$ to the center of $C_{n}^{\rm{A}}$ and from $A_q^{\rm{R}}$ to the center of $C_{n}^{\rm{Z}}$ onto the tunnel cross-section are defined as ${\theta _1} = {\arctan} \left(\frac{{{\tan}{\theta}_{p,n}^{\rm{T}}}}{{\left| {{\sin}{\phi}_{p,n}^{\rm{T}}} \right|}}\right)$ and ${\theta _2} = {\arctan} \left(\frac{{{\tan}{\theta}_{q,n}^{\rm{R}}}}{{\left| {{\sin}{\phi}_{q,n}^{\rm{R}}} \right|}}\right)$, respectively.			 
				The distance from $A_p^{\rm{T}}$ to the center of $C_{n}^{\rm{A}}$ is given as		
				\begin{equation}\label{D1}	
					D\!{_n^{\rm{T}}}\!(t) \!\!=\!\!
				 \left\{\!\! {\begin{array}{*{20}{c}}
							\!{\!\left| {\!y\!\! -\!\!\! {{A}}_{p,y}^{{\rm{T}}}(t)} \!\right|\!\sqrt {\!{{\!\left(\!{{\textstyle{1 \over {\!\cos\! {\theta _1}}}}} \!\right)\!}^2} \!\!\!+\! {\!{\left(\! {\tan \!{\phi}_{p,n}^{\rm{T}}} \!\right)\!}^2}\!,\!} }&{\!\!\!\!{\!\rm{if  }}\!\left| {\!y\!\! -\!\!\! {{A}}_{p,y}^{{\rm{T}}}(t)} \!\right| \!\!< \!\!\frac{{h \!-\! {{A}}_{\!p,z}^{{\!\rm{T}}}\!(t)}}{{\left| {\tan {\theta_1}} \right|}}}\\[10pt]
							{\!\!\frac{{h \!-\! {{A}}_{\!p,z}^{{\!\rm{T}}}\!(t)}}{{\left| {\tan {\theta_1}} \right|}}\!\sqrt {\!{{\!\left(\! {{\textstyle{1 \over {\!\cos {\!\theta _1}}}}} \!\right)\!}^2} \!\!\!+\! {\!{\left(\! {\tan \!{\phi}_{p,n}^{\rm{T}}} \!\right)\!}^2}\!,\!} }&{\!\!\!\!\!\!\!\!{\rm{if  }}\!\left| {\!y\!\! -\!\!\! {{A}}_{p,y}^{{\rm{T}}}(t)} \!\right|\!\! >\!\!\frac{{h \!-\! {{A}}_{\!p,z}^{{\!\rm{T}}}\!(t)}}{{\left| {\tan {\theta_1}} \right|}}}
					\end{array}} \right.
				\end{equation}where $y = 0$ and $y = w$ correspond to the left and right tunnel sidewalls, ${{A}}_{p,y}^{\rm{T}}(t)$ and ${{A}}_{p,z}^{\rm{T}}(t)$ represent the $y$- and $z$-components of ${\bf{A}}_p^{\rm{T}}(t)$, respectively.			
				Similarly, the distance from ${\bf{A}}_q^{\rm{R}}(t)$ to the center of cluster $C_n^{\rm{Z}}$ can be obtained by applying (\ref{D1}), with ${\theta_1}$, ${\phi}_{p,n}^{\rm{T}}$, ${\theta}_{p,n}^{\rm{T}}$, ${{A}}_{p,y}^{\rm{T}}(t)$, and ${{A}}_{p,z}^{\rm{T}}(t)$ replaced by ${\theta_2}$, ${\phi}_{q,n}^{\rm{R}}$, ${\theta}_{q,n}^{\rm{R}}$, ${{A}}_{q,y}^{\rm{R}}(t)$, and ${{A}}_{q,z}^{\rm{R}}(t)$, respectively.
				Here, ${{A}}_{q,y}^{{\rm{R}}}(t)$ and  ${{A}}_{q,z}^{{\rm{R}}}(t)$ denote the $y$- and \mbox{$z$-components} of ${\bf{A}}_q^{{\rm{R}}}(t)$.
				Based on the angular and distance parameters, the 3D coordinates of $C_{n}^{\rm{A}}$ and $C_{n}^{\rm{Z}}$ are shown as	
				\begin{equation}\label{}	
				 	{\bf{C}}_n^{\rm{A}}(t) = {\bf{A}}_p^{\rm{T}}(t) + D_n^{\rm{T}}(t){\left[ {\begin{array}{*{20}{c}}
				 	{{\cos}{\phi}_{p,n}^{\rm{T}}{\cos}{\theta}_{p,n}^{\rm{T}}}\\
				 	{{\sin}{\phi}_{p,n}^{\rm{T}}{\cos}{\theta}_{p,n}^{\rm{T}}}\\
				 		{\sin {\theta}_{p,n}^{\rm{T}}}
				 		\end{array}} \right]^{\rm{T}}}
				\end{equation}
			 	\begin{equation}\label{}	
						{\bf{C}}_n^{\rm{Z}}(t) = {\bf{A}}_q^{\rm{R}}(t) + D_n^{\rm{R}}(t){\left[ {\begin{array}{*{20}{c}}
			 			{{\cos}{\phi}_{q,n}^{\rm{R}}{\cos}{\theta}_{q,n}^{\rm{R}}}\\
			 			{{\sin}{\phi}_{q,n}^{\rm{R}}{\cos}{\theta}_{q,n}^{\rm{R}}}\\
			 			{\sin {\theta}_{q,n}^{\rm{R}}}
			 	\end{array}} \right]^{\rm{T}}}.
			 	\end{equation}

		 		To model the evolution of clusters and thereby account for channel non-stationarity, a birth-death process is employed. 
		 		Unlike the VR mechanism updating clusters according to their explicit geometric visibility, the birth-death process statistically governs cluster survival and generation, while both mechanisms update the active cluster set to characterize channel non-stationarity.		 		
		 		Within the time  interval $\Delta t$,  taking into account both the dynamic changes of propagation environment and the movement of transceiver, the survival probability of clusters can be calculated as
				\begin{equation}\label{Ps}
				 	{P_{\rm{S}}}\!\left( \!{\Delta t} \right) \!\!=\! \!\exp \!\left({\!\! - \!\lambda _{\rm{R}}^{}\!\frac{{\!\Delta t \!\left( {\left\| \!{{{\bf{v}}_{\rm{R}}}\!(t)} \!\right\| \!+\! \left\|\! {{{\bf{v}}_{\rm{T}}}\!(t)}\! \right\| \!+\! \left\|\! {{\bf{v}}_n^{\rm{A}}(t)}\! \right\|\! +\! \left\|\! {{\bf{v}}_n^{\rm{Z}}(t)} \!\right\|}\! \right)}}{{D_c^{\rm{T}}}}}\!\right) 
				\end{equation}  
				where ${{D_c^{\rm{T}}}}$ represents the correlated distance coefficient in time domain. 		 
				Consequently, the expected number of newly generated clusters within $\Delta t$ can be determined by
				\begin{equation}\label{Nnew}
				\mathbb{E}\left[ {N_{{\rm{new}}}\left( {\Delta t} \right)} \right] = \frac{{\lambda _{\rm{G}}}}{{\lambda _{\rm{R}}}}\left[ {1 - P_{{\rm{S}}}\left( {\Delta t} \right)} \right].
				\end{equation}

			\subsubsection{Cave Scenario}
				In the cave scenario, both spatially dispersed clusters and directionally confined clusters exist due to the complex and irregular structure.
				The proportion of these two cluster types can be given as $\eta(t)=\frac{{N _{1}}(t)}{{N _{2}}(t)}$, where ${{N _{1}}(t)}$ and ${{N _{2}}(t)}$ denote the time-variant number of spatially dispersed clusters and directionally confined clusters, respectively.
				For spatially dispersed clusters generated according to PPP, the scattering cluster density ${\lambda_{\rm{P}}}$ is lower than that in the pillar scenario, reflecting a sparser scattering environment.
				Furthermore, the VR is also utilized to describe the evolution of spatially dispersed clusters in the cave scenario.
				For directionally confined clusters generated according to the angular spread parameters, the generation rate $\lambda_{\rm{G}}$ and recombination rate $\lambda_{\rm{R}}$ are lower than those in the tunnel scenario due to weaker structural confinement.
				Besides, the birth-death process is also applied to characterize the evolution of directionally confined clusters in the cave scenario.			
\vspace{-0.3cm} 
		 \subsection{Generation and Evolution of Ray Parameters}\label{DMCSMC}			 
			To further accurately depict the actual propagation environment in underground spaces, both SMCs and DMCs are incorporated into the proposed GBSM.		  
			The angles of DMCs are concentrated around the same angles as those of SMCs, but with a wider angular spread.			
			In the case of SMCs, the azimuth and elevation angles are represented as ${\phi}_{p(q),m_n}^{\rm{T(R)}} = {\phi}_{p(q),n}^{\rm{T(R)}} + \Delta \phi _{{m_n}}^{{\rm{T(R)}}}$ and ${\theta}_{p(q),m_n}^{\rm{T(R)}} = {\theta}_{p(q),n}^{\rm{T(R)}} + \Delta \theta _{{m_n}}^{{\rm{T(R)}}}$. 
			Similarly, ${\phi}_{p(q),l_n}^{\rm{T(R)}} = {\phi}_{p(q),n}^{\rm{T(R)}} + \Delta \phi _{{l_n}}^{{\rm{T(R)}}}$ and ${\theta}_{p(q),l_n}^{\rm{T(R)}} = {\theta}_{p(q),n}^{\rm{T(R)}} + \Delta \theta _{{l_n}}^{{\rm{T(R)}}}$ describe the azimuth and elevation angles of DMCs.
			The azimuth angle offsets, i.e., $\Delta \phi _{{m_n}}^{{\rm{T(R)}}}$ and $\Delta \phi _{{l_n}}^{{\rm{T(R)}}}$, and elevation angle offsets, i.e., $\Delta \theta _{{m_n}}^{{\rm{T(R)}}}$ and $\Delta \theta _{{l_n}}^{{\rm{T(R)}}}$, follow zero-mean Laplacian distributions with different standard deviations.			
			Furthermore, the standard deviations of $\Delta \phi _{{l_n}}^{{\rm{T(R)}}}$  and $\Delta \theta _{{l_n}}^{{\rm{T(R)}}}$ are larger than that of $\Delta \phi _{{m_n}}^{{\rm{T(R)}}}$  and $\Delta \theta _{{m_n}}^{{\rm{T(R)}}}$.		
			Due to the different spatial dispersions of scattering clusters, the distance offsets of rays associated with SMCs and DMCs relative to the cluster center, i.e., ${\Delta d_{{m_{n}}}^{\rm{A(Z)}}}$ and ${\Delta d_{{l_{n}}}^{\rm{A(Z)}}}$, are postulated to conform to Gaussian distributions.
			Therefore, the  distance vectors of SMC and DMC rays can be calculated as 
			\begin{equation}\label{}	
			{\bf{D}}_{{m_n}}^{\rm{A(Z)}}(t)\!=\! \left(D_n^{{\rm{T(R)}}}(t)\!+\!{\Delta d_{{m_n}}^{\rm{A(Z)}}}\right)\!{\left[\hspace{-0.2cm}{\begin{array}{*{20}{c}}
						{{\cos}{\phi}_{p(q),{{m_n}}}^{\rm{T(R)}}{\cos}{\theta}_{p,{{m_n}}}^{\rm{T(R)}}}\\
						{{\sin}{\phi}_{p(q),{{m_n}}}^{\rm{T(R)}}{\cos}{\theta}_{p,{{m_n}}}^{\rm{T(R)}}}\\
							{\sin {\theta}_{p(q),{{m_n}}}^{\rm{T(R)}}}
					\end{array}}\hspace{-0.2cm} \right]^{\rm{T}}}
			\end{equation}
			\begin{equation}\label{}	
				{\bf{D}}_{{l_n}}^{\rm{A(Z)}}(t) \!=  \!\left(D_n^{{\rm{T(R)}}}(t) \!+\!{\Delta d_{{l_n}}^{\rm{A(Z)}}}\right)\!{\left[ \hspace{-0.2cm}{
					\begin{array}{*{20}{c}}{{\cos}{\phi}_{p(q),{{l_n}}}^{\rm{T(R)}}{\rm{cos}}{\theta}_{p(q),{{l_n}}}^{\rm{T(R)}}}\\
						{{\sin}{\phi}_{p(q),{{l_n}}}^{\rm{T(R)}}{\rm{cos}}{\theta}_{p(q),{{l_n}}}^{\rm{T(R)}}}\\
							{\sin {\theta}_{p(q),{{l_n}}}^{\rm{T(R)}}}
					\end{array}}\hspace{-0.2cm} \right]^{\rm{T}}}.
			\end{equation}
		
			Based on geometry relationships, the delays of SMC and DMC rays can be calculated as 
			\begin{small}
				\begin{equation}\label{tau1}	
					\tau _{qp,{m_n}}(t) = \frac{{\left\| {{\bf{D}}_{{m_n}}^{\rm{A}}(t)} \right\| + \left\| {{\bf{D}}_{{m_n}}^{\rm{Z}}(t)} \right\| }}{c} + {{\tilde \tau }_{n}}
				\end{equation}
			\end{small}
			\begin{small}
				\begin{equation}\label{tau2}	
					\tau _{qp,{l_n}}(t) = \frac{{\left\| {{\bf{D}}_{{l_n}}^{\rm{A}}(t)} \right\| + \left\| {{\bf{D}}_{{l_n}}^{\rm{Z}}(t)} \right\| }}{c} + {{\tilde \tau }_{n}} + \tau _{{l_n}}^{{\rm{add}}}
				\end{equation}
			\end{small}where $\tau _{{l_n}}^{{\rm{add}}}$ is the additional random delay \cite{J25_TVT_LZhang_DMCgbsm}, 
			$\tilde \tau _{{n}} = \frac{{\left\| {{\bf{C}}_{{n}}^{\rm{A}}(t) - {\bf{C}}_{{n}}^{\rm{Z}}(t)} \right\|}}{c} - r_\tau \sigma _\tau {\rm{ln}}{u_n}$ is the delay caused by propagation distance of virtual link. 
			Here, $u_n \sim \mathcal{U}(0,1)$, $r_\tau$ denotes the delay scaling factor, and $\sigma_\tau$ represents the delay spread following a log-normal distribution.
			Furthermore, the powers of SMC and DMC rays can be calculated as 
				\begin{equation}\label{smcpower}
					\tilde P_{qp{\rm{,}}{m_n}}(t) = {e^{ - \frac{{\Delta \tau _{{m_n}}}}{{\sigma _{\tau}}}}}
				\end{equation}
				\begin{equation}\label{dmcpower}
					\tilde P_{qp{\rm{,}}{l_n}}(t) = {e^{ - {\beta _d}\Delta \tau _{{l_n}}}}
			\end{equation}where ${\Delta \tau _{{m_n}/{l_n}}}$ is the delay offset of each SMC/DMC ray with respect to the center of cluster, and ${\beta _d}$ is the power decay factor \cite{J19_AWPL_JLi_IndoorDMC}.	 
			After normalization, the powers of of SMC and DMC rays can be obtained as
			\begin{equation}\label{normalizedSMCpower}
				P_{qp,{m_n}}(t) = \frac{{\tilde P_{qp,{m_n}}(t)}}{{\sum\limits_{n = 1}^{N(t)}{\sum\limits_{{m_n} = 1}^{{M_n}(t)} {\tilde P_{qp,{m_n}}(t)} } }}
			\end{equation}
			\begin{equation}\label{normalizedDMCpower}
				P_{qp,{l_n}}(t) = \frac{{\tilde P_{qp,{l_n}}(t)}}{{\sum\limits_{n = 1}^{N(t)}{\sum\limits_{{l_n} = 1}^{{L_n}(t)} {\tilde P_{qp,{l_n}}(t)} } }}.
			\end{equation}
			
			During cluster evolution, the powers of SMC and DMC rays within a dying cluster decay gradually, whereas those within a newly birth cluster increase smoothly. To characterize the different power variation rates of SMC and DMC rays, a squared-cosine function parameterized by the roughness attenuation factor is adopted.		
	  		The power evolution of rays in dying and birth clusters are modeled as \cite{J25_TVT_LZhang_DMCgbsm}
	  		\begin{equation}\label{wd}
	  			w_{\mathrm{d}}(k) = \cos^2\left( \frac{\pi}{2 T_{\rm{BD}} \rho_{\mathrm{r}}} k \Delta t \right), \quad k \Delta t \in [0, T_{\rm{BD}} \rho_{\mathrm{r}}]
	  		\end{equation}
	  		\begin{equation}\label{wb}
	  			w_{\mathrm{b}}(k) = 1 - w_{\mathrm{d}}(k)
	  		\end{equation}
	  		where $T_{\rm{BD}}$ denotes the time interval  for cluster birth and death, and $\rho_{\mathrm{r}}$ is the roughness attenuation factor
	  		that can be represented as \cite{J17_TAP_CZhou_Surfaceroughness}
		    \begin{equation}\label{rho_r}
		    	{\rho _{\rm{r}}} = \exp \left\{ { - 2{{\left( {\frac{{2\pi \Delta s\cos {\phi} }}{\lambda }} \right)}^2}} \right\}.
		    \end{equation}A smaller ${\rho _{\rm{r}}}$ indicates a rougher surface, leading to faster power dynamics. 
	    	Here, $ {\phi} = {\phi}_{p,l_n}^{\rm{T}}$ for DMCs, and  $ {\phi} = {\phi}_{p,m_n}^{\rm{T}}$ for SMCs. 
	    	Since SMCs and DMCs follow distinct angular distributions, their ray-dependent values of $\rho_{\rm r}$ generally differ, resulting in different power transition rates~\cite{J25_TVT_LZhang_DMCgbsm}. Moreover, the transition duration $T_{\rm BD}\rho_{\rm r}$ depends on surface roughness, ray angle, and wavelength~\cite{J17_TAP_CZhou_Surfaceroughness}. 
	    	Therefore, the different underground structures yield different SMC/DMC power decay rates because of their distinct surface and scattering characteristics.		
    		In addition, the proposed model also explicitly incorporates the frequency-dependent additional attenuation, path phases/Doppler shifts, and SMC/DMC power transitions via (\ref{gam}), (\ref{h}), and (\ref{rho_r}), respectively.

	\section{Statistical Properties}\label{p3}
		\subsection{STF-CFs}
		    To model the correlation properties of multipath fading in space, time, and frequency domains, STF-CFs are given by  
			\begin{equation}\label{R}
				\begin{array}{l}
					{R_{qp,q'p'}}(t,f;\Delta {\xi _{\rm{T}}},\Delta {\xi _{\rm{R}}},\Delta t,\Delta f)\\[8pt]
					= \mathbb{E}\left\{ {{H_{qp}}(t,f){H_{q'p'}}^ * (t + \Delta t,f + \Delta f)} \right\}\\[8pt]
					=\frac{{{K_R}(t)}}{{{K_R}(t) + 1}}R_{qp,q'p'}^{{\rm{LoS}}}(t,f;\Delta {\xi _{\rm{T}}},\Delta {\xi _{\rm{R}}},\Delta t,\Delta f)\\[8pt]
					+ \frac{1}{{{K_R}(t) + 1}}
					[R_{qp,q'p'}^{{\rm{SMC}}}(t,f;\Delta {\xi _{\rm{T}}},\Delta {\xi _{\rm{R}}},\Delta t,\Delta f)\\[8pt]
					{\kern 1pt} {\kern 1pt} {\kern 1pt} {\kern 1pt} {\kern 1pt} {\kern 1pt} {\kern 1pt} {\kern 1pt} {\kern 1pt} {\kern 1pt} {\kern 1pt} {\kern 1pt}{\kern 1pt} {\kern 1pt} {\kern 1pt} {\kern 1pt} {\kern 1pt} {\kern 1pt}{\kern 1pt} {\kern 1pt} {\kern 1pt} {\kern 1pt} {\kern 1pt} {\kern 1pt}{\kern 1pt} {\kern 1pt} {\kern 1pt} {\kern 1pt} {\kern 1pt} {\kern 1pt}{\kern 1pt} {\kern 1pt} {\kern 1pt} {\kern 1pt} {\kern 1pt} {\kern 1pt}{\kern 1pt} {\kern 1pt} {\kern 1pt} {\kern 1pt} {\kern 1pt} {\kern 1pt}+  R_{qp,q'p'}^{{\rm{DMC}}}(t,f;\Delta {\xi _{\rm{T}}},\Delta {\xi _{\rm{R}}},\Delta t,\Delta f)]
				\end{array}
			\end{equation}where ${H_{qp}}\left( {t,f} \right)$ and ${H_{q'p'}}(t + \Delta t,f + \Delta f)$ are the time-varying channel transfer functions (CTFs) obtained by applying the Fourier transform on CIRs. 
			The correlations of the LoS component, SMC, and DMC are expressed as
			\begin{equation}\label{RLoS}
				\begin{array}{l}
				R_{qp,q'p'}^{{\rm{LoS}}}(t,f;\Delta {\xi _{\rm{T}}},\Delta {\xi _{\rm{R}}},\Delta t,\Delta f) \\[8pt]
				= {e^{j2\pi \Delta f\tau _{q'p'}^{{\rm{LoS}}}(t + \Delta t)}}{e^{j2\pi ({f_{\rm{c}}} - f)[\tau _{qp}^{{\rm{LoS}}}(t) - \tau_{q'p'}^{{\rm{LoS}}}(t + \Delta t)]}}
				\end{array}
			\end{equation}
			\begin{equation}\label{RSMC}
				\begin{array}{l}
					\begin{array}{*{20}{l}}
					R_{qp,q'p'}^{{\rm{SMC}}}(t,f;\Delta {\xi _{\rm{T}}},\Delta {\xi _{\rm{R}}},\Delta t,\Delta f)\\[8pt]
						{= {P_{\rm{S}}}\left( {\Delta t} \right) \cdot \mathbb{E}\left\{ {\sum\limits_{n = 1}^{N(t)}{\sum\limits_{{m_n} =1}^{{M_n}(t)} {\sqrt {{P_{qp,{m_n}}}(t){P_{q'p',{m_n}}}(t + \Delta t)} } } } \right.}
					\end{array}\\[8pt]
					\left. { \cdot {e^{j2\pi \Delta f{\tau_{q'p',{m_n}}}(t + \Delta t)}}{e^{j2\pi ({f_{\rm{c}}}- f)[{\tau _{qp,{m_n}}}(t) -{\tau _{q'p',{m_n}}}(t + \Delta t)]}}} \right\}\\[5pt]
				\end{array}
			\end{equation}
			\begin{equation}\label{RDMC}
				\begin{array}{l}
					\begin{array}{*{20}{l}}
						R_{qp,q'p'}^{{\rm{DMC}}}(t,f;\Delta {\xi _{\rm{T}}},\Delta{\xi _{\rm{R}}},\Delta t,\Delta f)\\[8pt]
						{ = {P_{\rm{S}}}\left( {\Delta t} \right) \cdot\mathbb{E}\left\{ {\sum\limits_{n =1}^{N(t)}{\sum\limits_{{l_n} = 1}^{{L_n}(t)} {\sqrt {{P_{qp,{l_n}}}(t){P_{q'p',{l_n}}}(t + \Delta t)} } } } \right.}
					\end{array}\\[8pt]
					\left. {{\kern 1pt} {\kern 1pt} {\kern 1pt} {\kern 1pt} {\kern 1pt} {\kern 1pt} {\kern 1pt} \cdot {e^{j2\pi \Delta f{\tau_{q'p',{l_n}}}(t + \Delta t)}}{e^{j2\pi ({f_{\rm{c}}} - f)[{\tau _{qp,{l_n}}}(t) - {\tau _{q'p',{l_n}}}(t + \Delta t)]}}} \right\}\\[5pt]
				\end{array}.
			\end{equation}Furthermore, the temporal autocorrelation function (TACF) ${R_{qp}}(t,f;\Delta t)$ can be derived by setting $q=q'$, $p=p'$, and $\Delta f=0$ in the STF-CF. The spatial cross-correlation function (SCCF) ${R_{qp,q'p'}}(t,f;\Delta {\xi _{\rm{T}}},\Delta {\xi _{\rm{R}}})$  is obtained by imposing $\Delta t=0$ and $\Delta f=0$ in the STF-CF. Similarly, the STF-CF can  reduce to the frequency correlation function (FCF) ${R_{qp}}(t,f;\Delta f)$ by setting $q=q'$, $p=p'$, and $\Delta t=0$. 
		\subsection{Delay PSD}
			The delay PSD characterizes how signal power is distributed across the delays of MPCs, and can be computed as
			\begin{equation}
				\begin{array}{l}
					{S_{qp}}\left( {t,\tau } \right) =  \!\sum\limits_{{n} = 1}^{N(t)} \!{\sum\limits_{m_{n} = 1}^{M_{n}(t)}(1\!-\!{\mu}(t)){P_{qp,m_{n}}(t)} } \delta \left( {\tau  -{\tau _{qp,m_{n}}}(t)} \right)\\[5pt]
					{\kern 1pt}{\kern 1pt}{\kern 1pt}{\kern 1pt}{\kern 1pt}{\kern 1pt}+\sum\limits_{{n} = 1}^{N(t)} {\sum\limits_{l_{n} = 1}^{L_{n}(t)}{\mu}(t){P_{qp,l_{n}}(t)} } \delta \left( {\tau  - {\tau _{qp,l_{n}}}(t)} \right)
					\\		
				\end{array}.
			\end{equation}The evolution of clusters is driven by changes in the channel environment, leading to time-varying delay PSD.
	
		\subsection{Angular PSD}
			The angular PSD represents the channel power distribution in angular domain. The angular PSD of AAoA~${\phi ^{\rm{R}}}$ and EAoA~${\theta^{\rm{R}}}$ can be expressed as
			\begin{equation}
				\begin{array}{l}
					{S_{qp}}\left( {t,{\phi^{\rm{R}}},{\theta^{\rm{R}}}} \right) = \\[5pt]
						\sum\limits_{{n} = 1}^{N(t)}\! {\sum\limits_{m_{n} = 1}^{M_{n}(t)}\!(\!1\!-\!{\mu}(t)\!)\!{P_{qp\!,m_{n}}(t)} } \delta \!\left(\! {{\phi^{\rm{R}}} \! -\! {\phi _{{q}\!,{m_{n}}}^{\rm{R}}}} \!\right)\!\delta \!\left( {\!{\theta^{\rm{R}}} \! -\! {\theta _{q\!,{m_{n}}}^{\rm{R}}}} \!\right)\\[5pt]
				\!+\!\sum\limits_{{n} = 1}^{N(t)} \!{\sum\limits_{l_{n} = 1}^{L_{n}(t)}\!{\mu}(t){P_{qp\!,l_{n}}(t)} } \delta \left( {{\phi^{\rm{R}}}  - {\phi _{q\!,{l_{n}}}^{\rm{R}}}} \right)\delta \left( {{\theta^{\rm{R}}}  \!-\! {\theta_{q\!,{l_{n}}}^{\rm{R}}}} \right)								
				\end{array}.
			\end{equation}Similarly, the angular PSD of AAoD and EAoD can be derived following the same way with ${\phi ^{\rm{R}}}$ and ${\theta^{\rm{R}}}$ replaced by ${\phi ^{\rm{T}}}$ and~${\theta^{\rm{T}}}$, respectively.
		\subsection{Doppler PSD}
			The Doppler frequency shift arises from relative motion among the Tx, Rx and scatterers, resulting in an offset of the carrier frequency. 
			To quantify the power distribution in the Doppler
			frequency, the Doppler PSD is expressed as
			\begin{equation}
				{S_{qp}}(t,f;\nu ) = \int_{ - \infty }^\infty  {{R_{qp}}(t,f;\Delta t)} {e^{ - j2\pi \nu \Delta t}}d\Delta t
			\end{equation}
			where $\nu$ denotes the Doppler frequency, and ${{R_{qp}}(t,f;\Delta t)}$ is the TACF  obtained from (\ref{R}).
		\subsection{RMS Delay Spread}
			The RMS delay spread serves as a key metric for assessing the temporal dispersion induced by multipath propagation. 
			A larger delay spread indicates more serious temporal dispersion, which can lead to increased inter-symbol interference.
			The RMS delay spread is given as
			\begin{equation}
				{\sigma _\tau } = \sqrt {\overline {{{\left( {\tau _{qp}^{}(t)} \right)}^2}}  - {\left({\overline { {\tau _{qp}^{}(t)} }}\right)^2}} 
			\end{equation}	
			where
			\begin{equation}
				\begin{array}{l}
					\overline {{{\left( {\tau _{qp}^{}(t)} \right)}^2}}  = \sum\limits_{n = 1}^{N(t)} {\sum\limits_{{m_n} = 1}^{{M_n}(t)} {(1-{\mu}(t))P_{qp,{m_n}}(t)} } {\left( {\tau_{qp,{m_n}}(t)} \right)^2}\\[5pt]
					 {\kern 1pt}{\kern 1pt}{\kern 1pt}{\kern 1pt}{\kern 1pt}{\kern 1pt}{\kern 1pt}{\kern 1pt}{\kern 1pt}{\kern 1pt}{\kern 1pt}{\kern 1pt}+\sum\limits_{n = 1}^{N(t)} {\sum\limits_{{l_n} = 1}^{{L_n}(t)}{{\mu}(t)P_{qp,{l_n}}(t)} } {\left( {\tau _{qp,{l_n}}(t)} \right)^2}						 			
				\end{array}
			\end{equation}	
			\begin{equation}
				\begin{array}{l}
					\overline { {\tau _{qp}^{}(t)} }  = 	 \sum\limits_{n = 1}^{N(t)} {\sum\limits_{{m_n} = 1}^{{M_n}(t)} {(1-{\mu}(t))P_{qp,{m_n}}(t)} } \tau _{qp,{m_n}}(t)\\[5pt]
					{\kern 1pt}{\kern 1pt}{\kern 1pt}{\kern 1pt}{\kern 1pt}{\kern 1pt}{\kern 1pt}{\kern 1pt}{\kern 1pt}{\kern 1pt}{\kern 1pt}{\kern 1pt}+\sum\limits_{n = 1}^{N(t)} {\sum\limits_{{l_n} = 1}^{{L_n}(t)} {{\mu}(t)P_{qp,{l_n}}(t)} } \tau _{qp,{l_n}}(t)				
				\end{array}.
			\end{equation}

		\subsection{Stationary Interval}	
			The stationary interval characterizes the temporal stability of fading statistics by representing the longest duration over which channel characteristics remain approximately invariant. 	
			Following the local stationarity criterion \cite{J23_TCOM_CXWang_PervasiveCM2}, the stationary interval is generally defined as the maximum time interval over which the temporal correlation of the delay PSD remains above a predefined threshold.
			Specifically, for a given threshold $c_{\rm thresh} \in (0,1)$, the stationary interval is represented by
			\begin{equation}
				{T_{{\rm{SI}}}}(t) = \max \{ \Delta t\left| {c(t,\Delta t)} \right. \ge {c_{{\rm{thresh}}}}\} 
			\end{equation}
			where ${c(t,\Delta t)}$ is the correlation coefficient that can be obtained by
			\begin{equation}
				c(t,\Delta t) = \frac{{\int {{S_{qp}}\left( {t,\tau } \right){S_{qp}}\left( {t + \Delta t,\tau } \right)d\tau } }}{{\max \left\{ {\int {{S_{qp}}{{\left( {t,\tau } \right)}^2}d\tau ,\int {{S_{qp}}{{\left( {t + \Delta t,\tau } \right)}^2}d\tau } } } \right\}}}.
		    \end{equation}
	
	\section{Results and Analysis}\label{p4}
		Simulated statistical properties of the proposed channel model are presented and discussed in this section.		
		The dimensions of the pillar, tunnel, and cave scenarios are defined based on  structural features.  
		For physical features, the surface roughness $\Delta s$ and wall tilt $\theta$ are set according to~\cite{J07_TAP_MRak_TunnelCave}, while the blockage probability  $P_{\rm{bloc}}$  is set to 0 with simulated navigation routes assumed free of temporary obstruction by humans or large machinery. This setting allows us to focus on channel characteristics under normal unblocked communication conditions.
		At cluster level, the scattering cluster density ${\lambda_{\rm{P}}}$ and VR $R_{\rm  lim}$ of spatially dispersed clusters are adopted from~\cite{J23_TAP_HChang_TunnelPillarGBSM}, the generation rate $\lambda_{\rm{G}}$  and recombination rate $\lambda_{\rm{R}}$ of directionally confined clusters are referenced from~\cite{J14_TCOM_AEForooshani_Tunnelangular}, and the cluster type proportion $\eta$ is set to 3/7 for the cave scenario. 
		Furthermore, the DMC power ratio for different underground scenaisrios is set according to the measurement results \cite{C20_EuCAP_SJiang_UndergroundDMCmea}.
		For clarity, Table~\ref{para_sim} provides a detailed listing of the parameters specific to different underground scenarios. 
		Unless otherwise specified, the simulation parameters are set as follows: $f_{\rm{c}}$~=~2.5~GHz, \mbox{$M_{\rm{T}}$ = $M_{\rm{R}}$ = 2}, \mbox{${v}_{\rm{R}}$ = 5 m/s}, \mbox{${v}_{\rm{T}}$ = 8 m/s}. 
		For simulations at other carrier frequencies, associated model parameters should be recalibrated based on corresponding channel measurements.		
		All scenarios adopt the same antenna configurations and Tx/Rx motion. 
		The AAV serves as the Tx and the AGV as the Rx, and they travel in opposite directions, first approaching and then receding from each other.

\begin{figure}[!t]			
			\centering	{\includegraphics[width=3.5in]{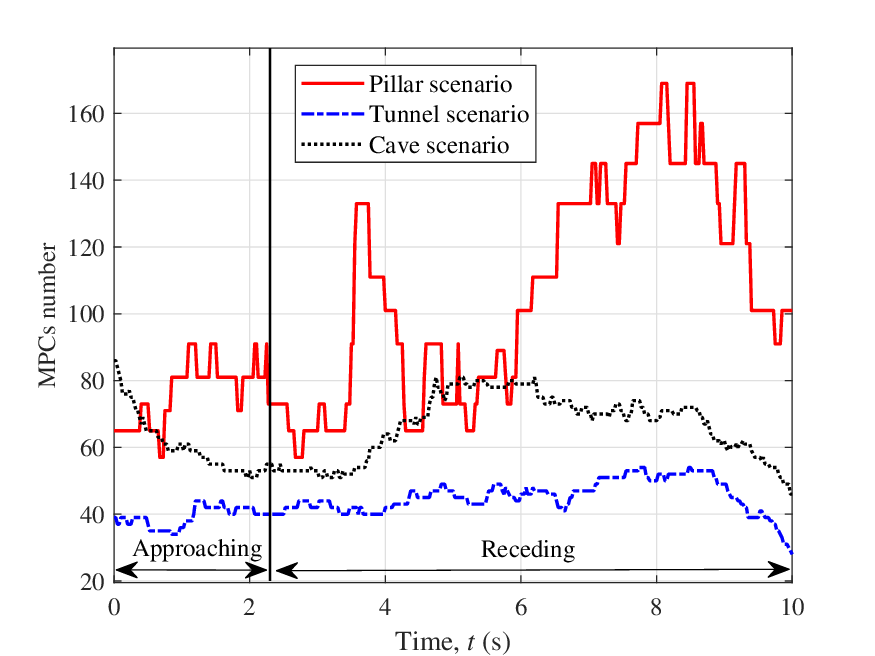}}
			\caption{MPCs numbers in different underground scenarios (\mbox{$f_{\rm{c}}$ = 2.5 GHz},  ${v}_{\rm{R}}$ = 5 m/s, ${v}_{\rm{T}}$~=~8~m/s, \mbox{$\alpha _{\rm{T(R)}} = \pi/3$}, \mbox{$\beta _{\rm{T(R)}} = \pi/4$}, \mbox{ $\gamma(t)=\beta(t) = \pi/4$}, $\varphi(t)=0$).}
			\label{Clusternumber}
		\end{figure}

		Fig.~\ref{Clusternumber} visualizes the MPC number in different underground scenarios. 
		In the pillar scenario, the large and fluctuating number of MPCs results from abundant and unevenly distributed scatterers. 
		The number of MPCs in the tunnel scenario fluctuates slightly and gradually decreases with distance, owing to the waveguide effect and the transition from near-field to far-field propagation.		
		In the cave scenario, the moderate MPC number and variability align with the weakened waveguide effect. 
		The consistency between scenario parameters, cluster evolution, and propagation characteristics verifies the rationality of the generation and evolution mechanisms of clusters and rays in different underground scenarios.

		\begin{table}[!t]
			\centering
			\caption{\small Parameters For Different Underground Scenarios.}
			\label{para_sim}
			\begin{tabular}{|m{1.8cm}<{\centering}|m{1.8cm}<{\centering}|m{1.8cm}<{\centering}|m{1.8cm}<{\centering}|}
				\hline
				\textbf{Parameter} & \textbf{Pillar} & \textbf{Tunnel} & \textbf{Cave} \\
				\hline
				$l$ & 150 m& 800 m& 400 m\\ \hline
				$w$  & 40 m& 10 m& 15 m\\ \hline
				$h$ & 4 m& 4 m& 6 m\\ \hline
				$\Delta s$ & 0.1 m& 0.01 m & 0.2 m\\ \hline
				$\theta$ & 1° & 0.01°& 5°\\ \hline
				${P_{{\rm{bloc}}}}$& 0 & 0& 0\\ \hline	 
				${\lambda_{\rm{P}}}$ & 0.004 m$^{-3}$& - & 0.002 m$^{-3}$\\ \hline
				$R_{\rm  lim}$& 20 m& -& 20 m\\ \hline
				$\lambda_{\rm{G}}$ & - & 20 & 20 \\  \hline
				$\lambda_{\rm{R}}$ & - & \shortstack{Near-field: 1 \\ Far-field: 4}& 4 \\ \hline			
				$\eta$ & - & -& 3/7 \\ \hline
				$\mu$  &  0.35 & 0.3 & 0.4  \\ \hline					
			\end{tabular}
		\end{table}		

\begin{figure}[!t]
			\centering
			\begin{subfigure}[!t]{0.45\textwidth}				
				\centering
				\includegraphics[width=1\textwidth]{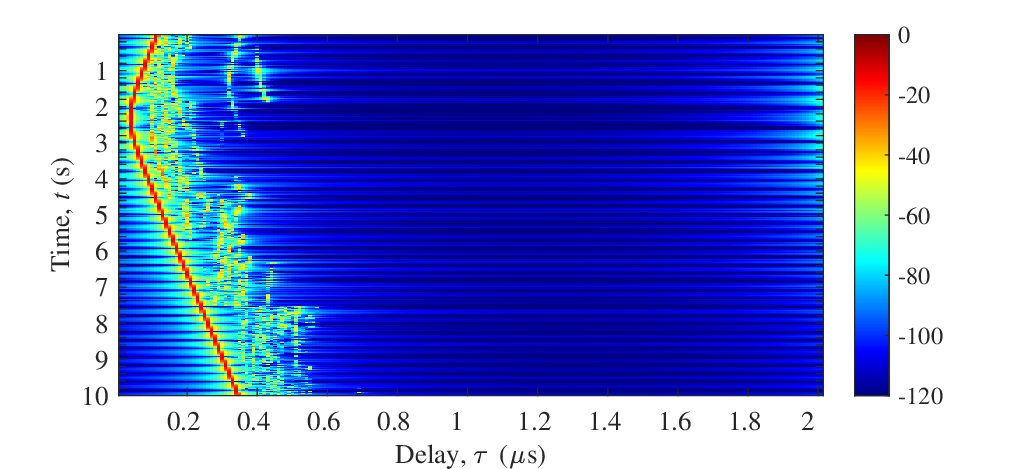}
				\caption{}
				\label{DelayPSDPillar}
			\end{subfigure}
			\hfill
			\begin{subfigure}[!t]{0.45\textwidth}
				\centering
				\includegraphics[width=1\textwidth]{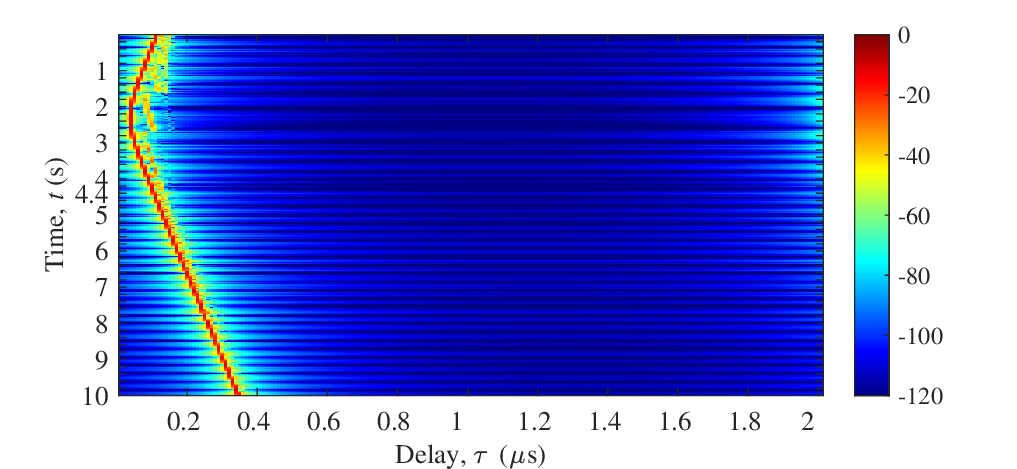}
				\caption{}
				\label{DelayPSDTunnel}
			\end{subfigure}	  
			\hfill
			\begin{subfigure}[!t]{0.45\textwidth}
				\centering
				\includegraphics[width=1\textwidth]{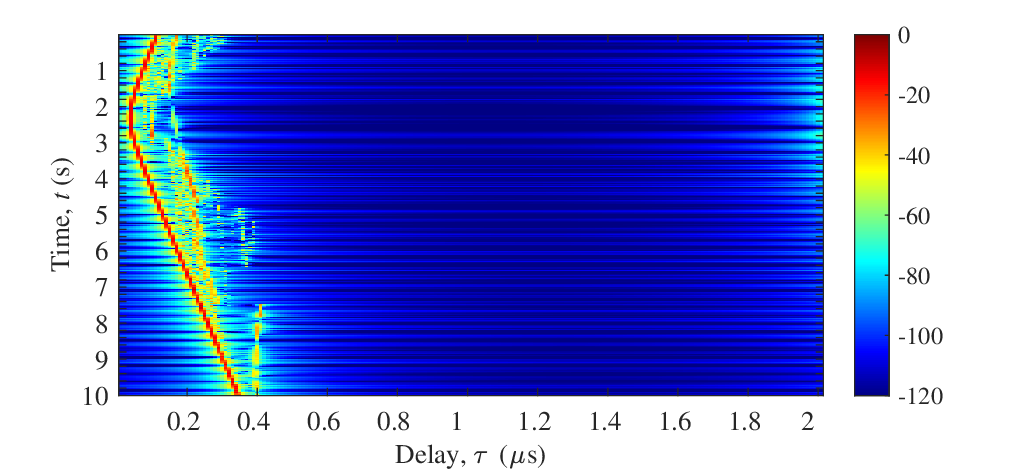}
				\caption{}
				\label{DelayPSDCave}
			\end{subfigure}
			\caption{Normalized time-variant delay PSDs of the proposed channel model in different cases: (a) Pillar scenario, (b) Tunnel scenario, and (c) Cave scenario (\mbox{$f_{\rm{c}}$ = 2.5 GHz}, \mbox{${v}_{\rm{R}}$ = 5 m/s}, \mbox{${v}_{\rm{T}}$ = 8 m/s}, \mbox{$\alpha _{\rm{T(R)}} = \pi/3$}, \mbox{$\beta _{\rm{T(R)}} = \pi/4$}, \mbox{ $\gamma(t)=\beta(t) = \pi/4$}, $\varphi(t)=0$).}
			\label{DelayPSD}
		\end{figure}

\begin{figure}[!t]
		   	\centering
		   	\begin{subfigure}[!t]{0.45\textwidth}
		   		\centering
		   		\includegraphics[width=1\textwidth]{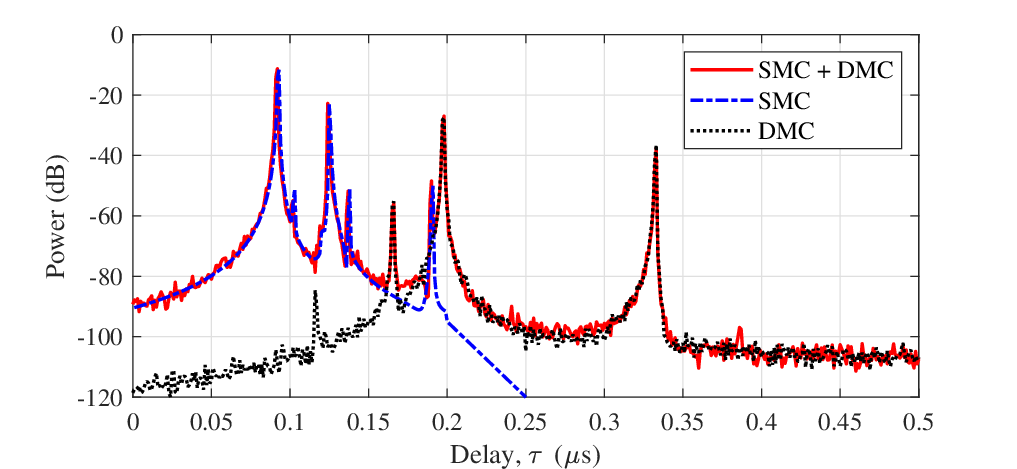}
		   		\caption{}
		   		\label{PDPPillar}
		   	\end{subfigure}   	
		   	\hfill
		   	\begin{subfigure}[!t]{0.45\textwidth}
		   		\centering
		   		\includegraphics[width=1\textwidth]{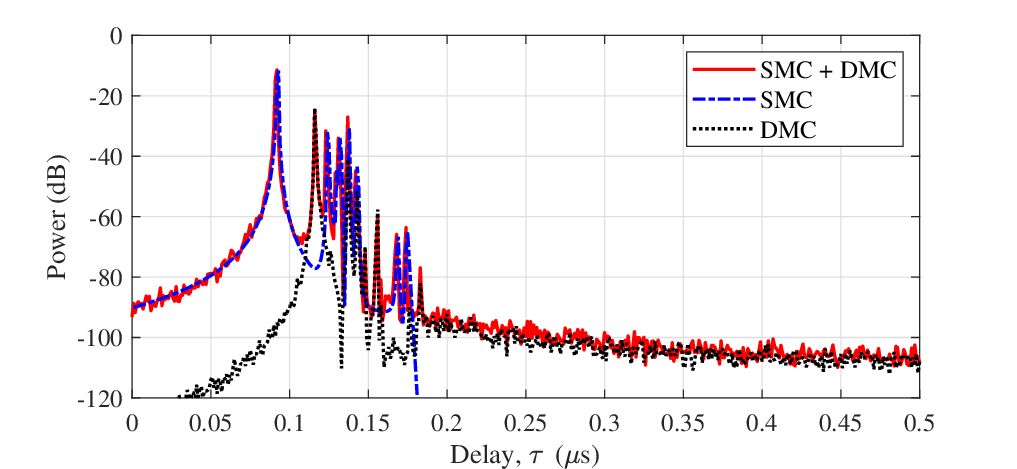}
		   		\caption{}
		   		\label{PDPTunnel}
		   	\end{subfigure}	  
		   	\hfill
		   	\begin{subfigure}[!t]{0.45\textwidth}
		   		\centering
		   		\includegraphics[width=1\textwidth]{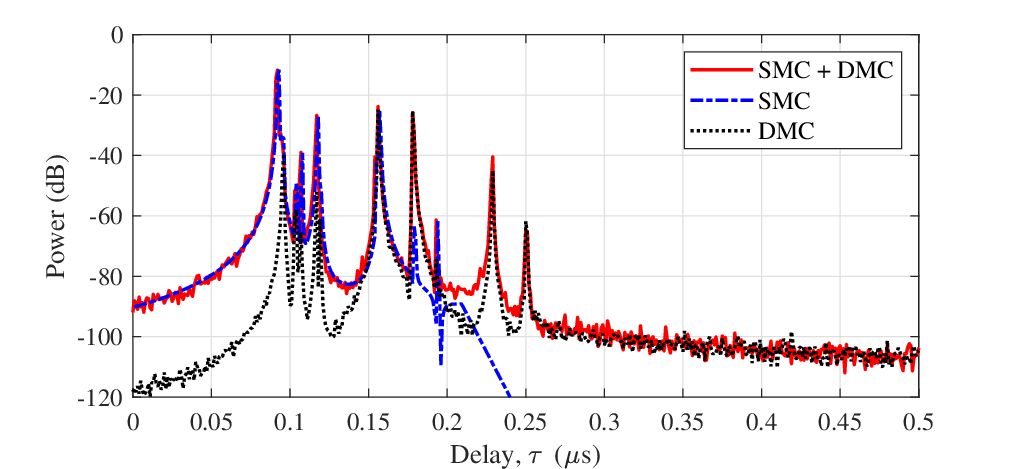}
		   		\caption{}
		   		\label{PDPCave}
		   	\end{subfigure}
		   	\caption{Delay PSDs at $t$ = 0.3 s for different cases: (a) Pillar scenario, (b) Tunnel scenario, and (c) Cave scenario (\mbox{$f_{\rm{c}}$ = 2.5 GHz}, \mbox{${v}_{\rm{R}}$ = 5 m/s}, \mbox{${v}_{\rm{T}}$ = 8 m/s}, \mbox{$\alpha _{\rm{T(R)}} = \pi/3$}, \mbox{$\beta _{\rm{T(R)}} = \pi/4$}, \mbox{ $\gamma(t)=\beta(t) = \pi/4$}, $\varphi(t)=0$). }
		   	\label{PDP}
		\end{figure}

		Fig.~\ref{DelayPSD} illustrates the normalized time-variant delay PSDs in different underground scenarios.
		The delays and powers of MPCs drift along the trajectory over time, accompanied by the birth and death  of MPCs.
		In the pillar scenario, abundant scatterers results in a pronounced delay power tail and a gradual attenuation of power. Conversely, the tunnel scenario is characterized by a relatively concentrated delay power distribution accompanied by rapid attenuation, which can be attributed to the waveguide effect.
		The cave scenario demonstrates intermediate characteristics in both delay power distribution and attenuation rate, attributable to moderate scattering conditions.
		Furthermore, Fig. \ref{PDP} shows the delay PSDs at $t$ = 0.3 s in different  different underground scenarios with and without DMCs. 
		It can be found that DMCs can contribute to the weaker parts of the delay PSD by filling in the power between or near the peaks of SMCs, and form an exponentially decaying tail with increasing delay.
		These results underscore the critical contribution of DMCs to accurate characterizing of underground spaces.
		
\begin{figure}[!t]
		 	\centering	{\includegraphics[width=3.5in]{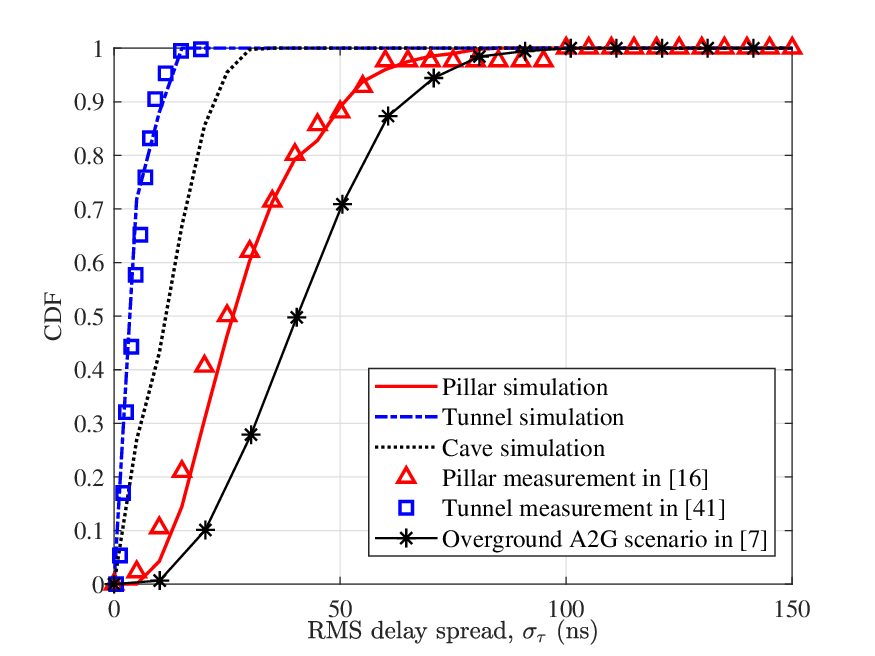}}
		 	\caption{CDFs of RMS delay spreads for the proposed channel model in different underground scenarios, measurement data of pillar scenario in~\cite{J23_TAP_HChang_TunnelPillarGBSM}, and measurement data of tunnel scenario in~\cite{J08_TAP_MBoutin_TunnelmeaDScom} (\mbox{$f_{\rm{c}}$ = 2.5 GHz}, \mbox{${v}_{\rm{R}}$ = 5 m/s}, \mbox{${v}_{\rm{T}}$ = 8 m/s}, \mbox{$\alpha _{\rm{T(R)}} = \pi/3$}, \mbox{$\beta _{\rm{T(R)}} = \pi/4$}, \mbox{ $\gamma(t)=\beta(t) = \pi/4$}, $\varphi(t)=0$).}
		 	\label{RMSDS}
		 \end{figure}

	    Fig.~\ref{RMSDS} compares the RMS delay spread in different underground scenarios.
	    The delay spread increases progressively from the tunnel to the cave, and further to the pillar scenario.	    
	    The simulation results are validated by comparing with available channel measurements~\cite{J23_TAP_HChang_TunnelPillarGBSM,J08_TAP_MBoutin_TunnelmeaDScom}.
	    In~\cite{J23_TAP_HChang_TunnelPillarGBSM}, channel measurements were conducted at 2.5~GHz in an underground garage. Although in~\cite{J08_TAP_MBoutin_TunnelmeaDScom}, channel measurements were conducted at 2.4~GHz in underground mine tunnels, the close proximity between 2.4~GHz and 2.5~GHz allows their results  to be sufficiently used for comparative analysis at 2.5~GHz.
	    Furthermore, for a fair comparison under consistent conditions, the RMS delay spread obtained from the overground air-to-ground (A2G) channel model~\cite{J21_IOT_HChang_UAVGBSM} is evaluated using identical antenna configuration and Tx/Rx movement.       
  		The RMS delay spread of the autonomous vehicle channel model in overground space is larger than that in underground space.
		The larger RMS delay spread of the overground A2G channel model results from the more widely distributed scatterers and larger path-length differences in the open environment. In contrast, confined underground structures restrict the spatial distribution and delays of MPCs.
	    
\begin{figure}[!t]
		 	\centering	{\includegraphics[width=3.5in]{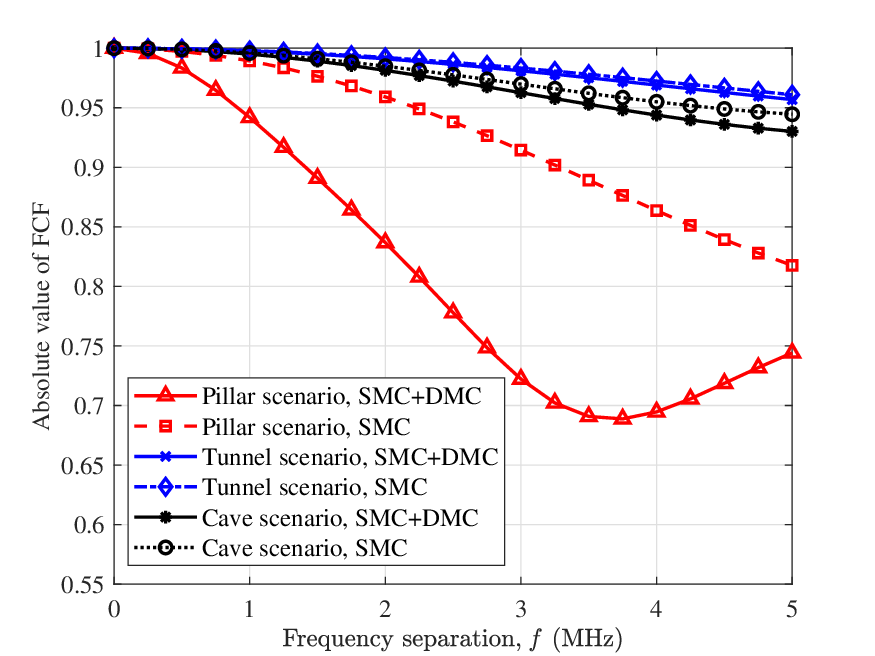}}
		 	\caption{FCFs of the proposed channel model in different underground scenarios (\mbox{$f_{\rm{c}}$ = 2.5 GHz}, \mbox{${v}_{\rm{R}}$ = 5 m/s}, \mbox{${v}_{\rm{T}}$ = 8 m/s}, \mbox{$\alpha _{\rm{T(R)}} = \pi/3$}, \mbox{$\beta _{\rm{T(R)}} = \pi/4$}, \mbox{ $\gamma(t)=\beta(t) = \pi/4$}, $\varphi(t)=0$).}
		 	\label{FCF}
		\end{figure}

	    Fig.~\ref{FCF} presents the FCFs in different underground scenarios. 
	    The absolute values of FCFs vary with frequency separation, reflecting the channel variation in frequency domain.
	    In the pillar scenario, abundant scatterers produce MPCs that contribute to a large delay spread,  leading to rapid FCF decay and significant frequency-selective fading.	    
	    In contrast, the waveguide effect concentrates the MPCs in the tunnel scenario, leading to relatively flat fading, thereby minimizing FCF decay.
	    Moreover, the results demonstrate that neglecting DMCs overestimates the FCF, resulting in overly optimistic performance predictions.
	    
\begin{figure}[!t]
	    	\centering
	    	\begin{subfigure}[!t]{0.48\textwidth}
	    		\centering
	    		\includegraphics[width=1\textwidth]{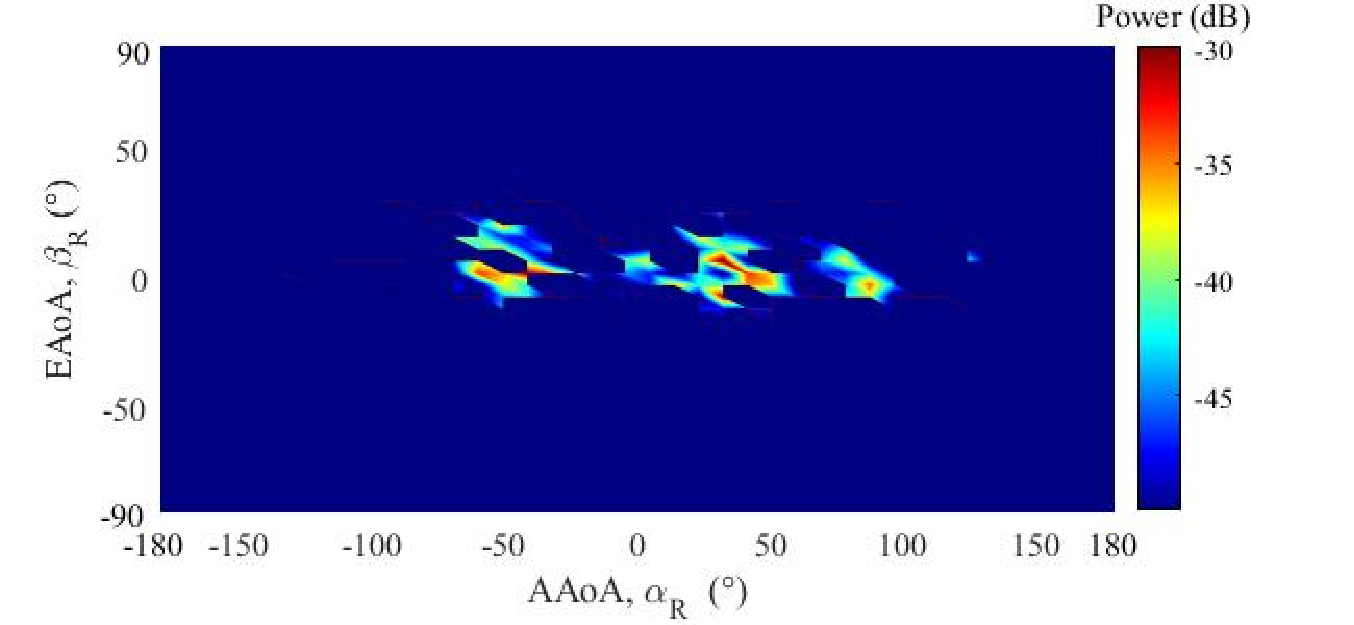}
	    		\caption{}
	    		\label{Pillar}
	    	\end{subfigure}
	    	\hfill
	    	\begin{subfigure}[!t]{0.48\textwidth}
	    		\centering
	    		\includegraphics[width=1\textwidth]{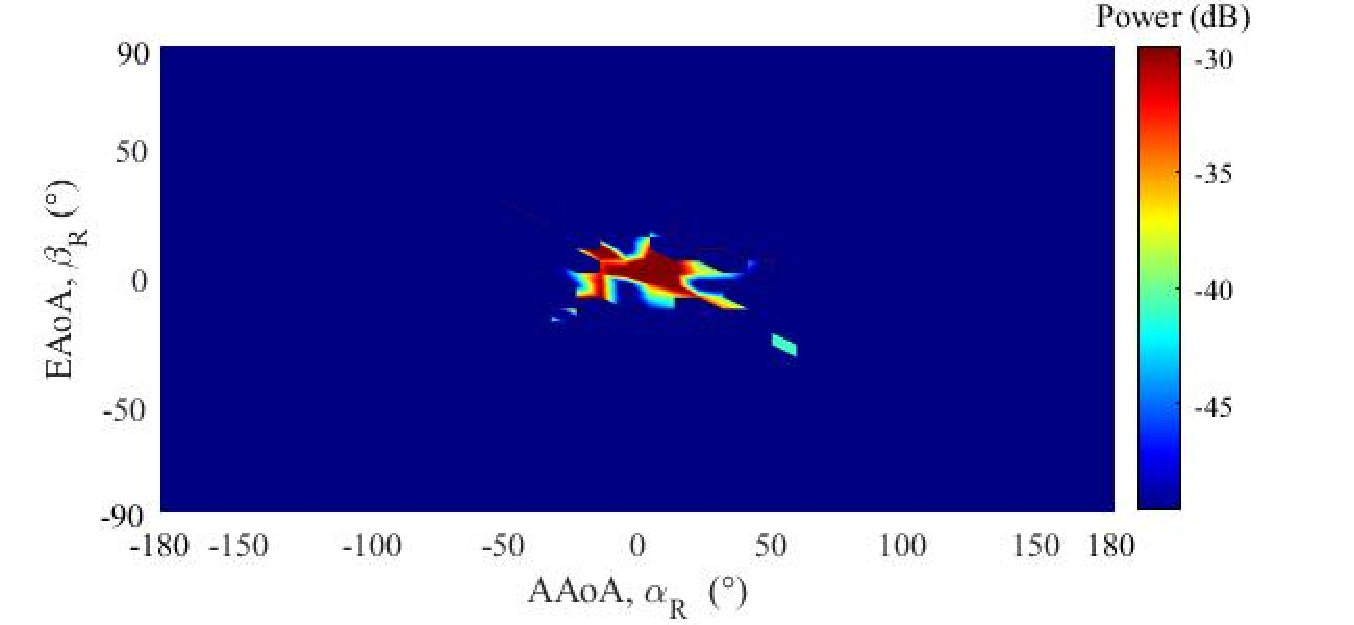}
	    		\caption{}
	    		\label{Tunnel}
	    	\end{subfigure}	  
	    	\hfill
	    	\begin{subfigure}[!t]{0.48\textwidth}
	    		\centering
	    		\includegraphics[width=1\textwidth]{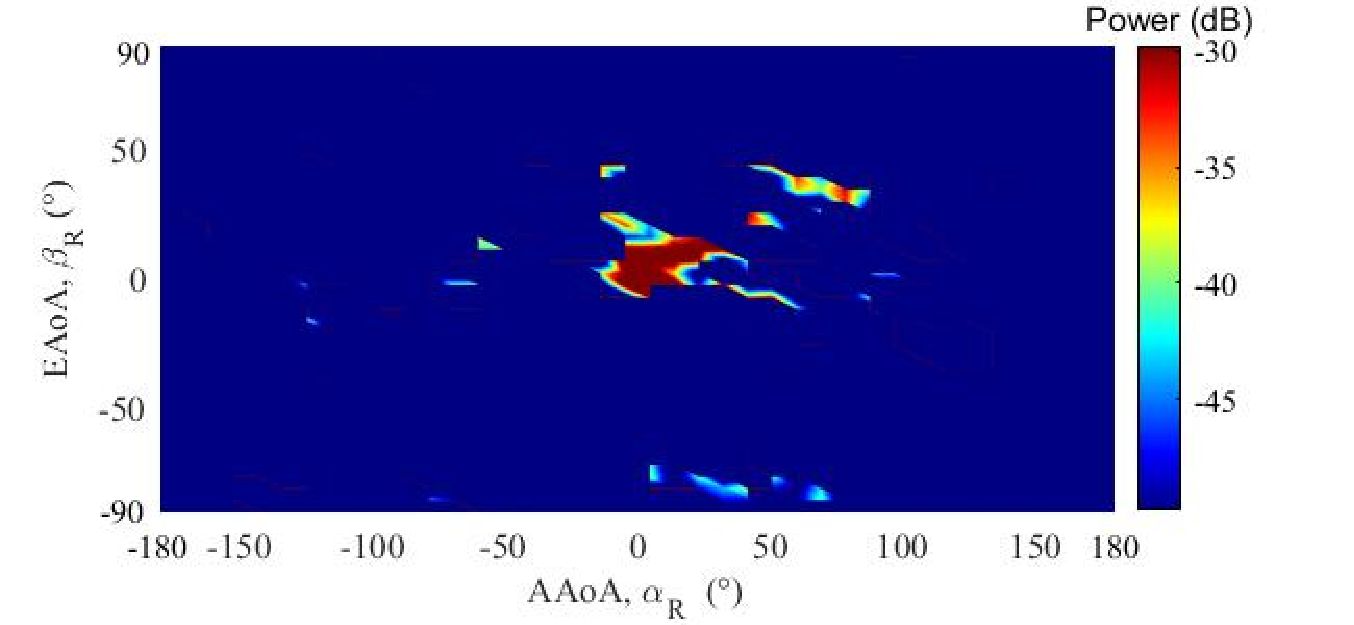}
	    		\caption{}
	    		\label{Cave}
	    	\end{subfigure}
	    	\caption{Angular PSDs of the proposed channel model in different cases: (a) Pillar scenario, (b) Tunnel scenario, and (c) Cave scenario (\mbox{$f_{\rm{c}}$ = 2.5 GHz}, \mbox{${v}_{\rm{R}}$ = 5 m/s}, \mbox{${v}_{\rm{T}}$ = 8 m/s}, \mbox{$\alpha _{\rm{T(R)}} = \pi/3$}, \mbox{$\beta _{\rm{T(R)}} = \pi/4$}, \mbox{ $\gamma(t)=\beta(t) = \pi/4$}, $\varphi(t)=0$).}
	    	\label{AnglePSD}
	    \end{figure}

\begin{figure}[!t]
			\centering	{\includegraphics[width=3.5in]{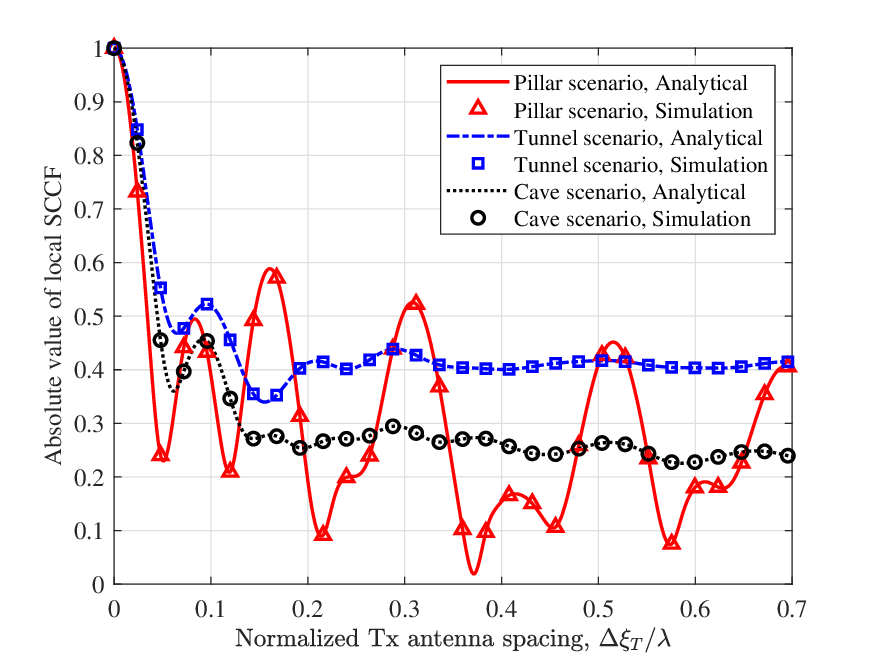}}
			\caption{SCCFs of the proposed channel model in different underground scenarios (\mbox{$f_{\rm{c}}$ = 2.5 GHz}, \mbox{${v}_{\rm{R}}$ = 5 m/s}, \mbox{${v}_{\rm{T}}$ = 8 m/s}, \mbox{$\alpha _{\rm{T(R)}} = \pi/3$}, \mbox{$\beta _{\rm{T(R)}} = \pi/4$}, \mbox{ $\gamma(t)=\beta(t) = \pi/4$}, $\varphi(t)=0$).}
			\label{SCCF}
		\end{figure}

	    Fig.~\ref{AnglePSD} illustrates angular PSDs in the pillar, tunnel, and cave scenarios.	
	    With vertical limitations from the floor and ceiling, angular dispersion in the azimuth direction tends to be typically dominant relative to that in the elevation direction.
	    Differences in angular and power distributions among these scenarios stem from the structural and physical features.
		Fig.~\ref{SCCF} shows the SCCFs in the pillar, tunnel, and cave scenarios.
		The SCCF in the tunnel scenario exhibits the least fluctuation and slowest decay, whereas that in the pillar scenario shows the most pronounced fluctuation and fastest decay.
		This behavior arises from differences in the angular spread of MPCs.
		In the tunnel scenario, the confined waveguide structure results in a narrow angular spread, leading to small phase shifts and thus high spatial correlation.
		In contrast, the pillar and cave scenarios exhibit larger angular spreads, which leads to faster SCCF decay.

\begin{figure}[!t]
			\centering	{\includegraphics[width=3.5in]{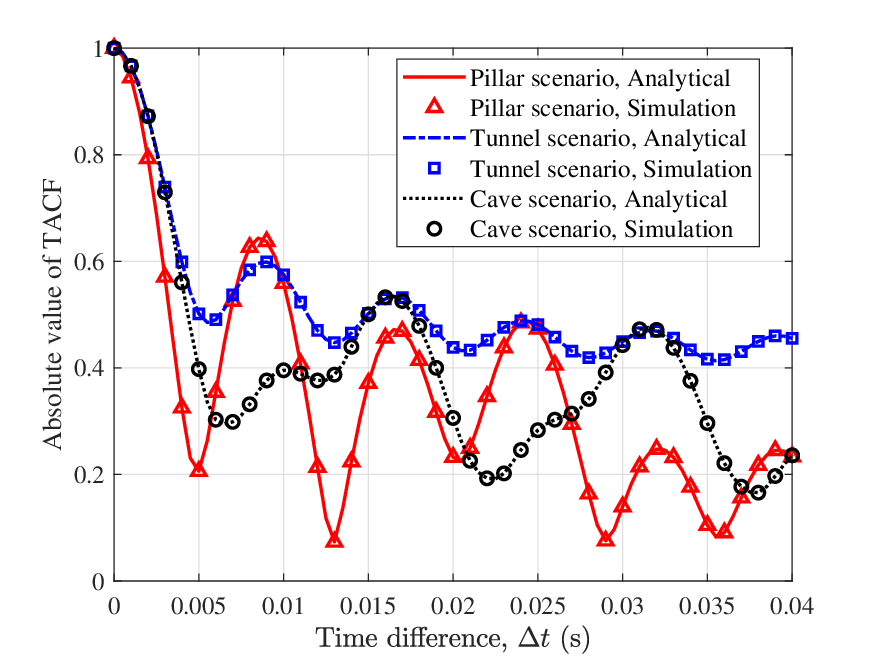}}
			\caption{TACFs of the proposed channel model in different underground scenarios (\mbox{$f_{\rm{c}}$ = 2.5 GHz}, \mbox{${v}_{\rm{R}}$ = 5 m/s}, \mbox{${v}_{\rm{T}}$ = 8 m/s}, \mbox{$\alpha _{\rm{T(R)}} = \pi/3$}, \mbox{$\beta _{\rm{T(R)}} = \pi/4$}, \mbox{ $\gamma(t)=\beta(t) = \pi/4$}, $\varphi(t)=0$).}
			\label{TACF}
		\end{figure}

		Fig.~\ref{TACF} shows the TACFs in different underground scenarios. 
		It is observed that the TACF curves exhibit trends similar to those of the SCCF.
		Under the same Tx/Rx mobility, the variation trends of the TACF in different underground scenarios are primarily determined by the spatial distribution and birth-death processes of clusters.	
		In the pillar scenario, spatially dispersed clusters and highly fluctuating MPCs lead to the most pronounced TACF fluctuations.
		In contrast, tunnel scenario features directionally confined clusters and more consistent MPCs, resulting in the smallest TACF fluctuations.	
		In addition, the consistency between simulation results from CIR correlation calculations and analytical results from~(\ref{R}) confirms the validity of the model.
	
\begin{figure}[!t]
			\centering	{\includegraphics[width=3.5in]{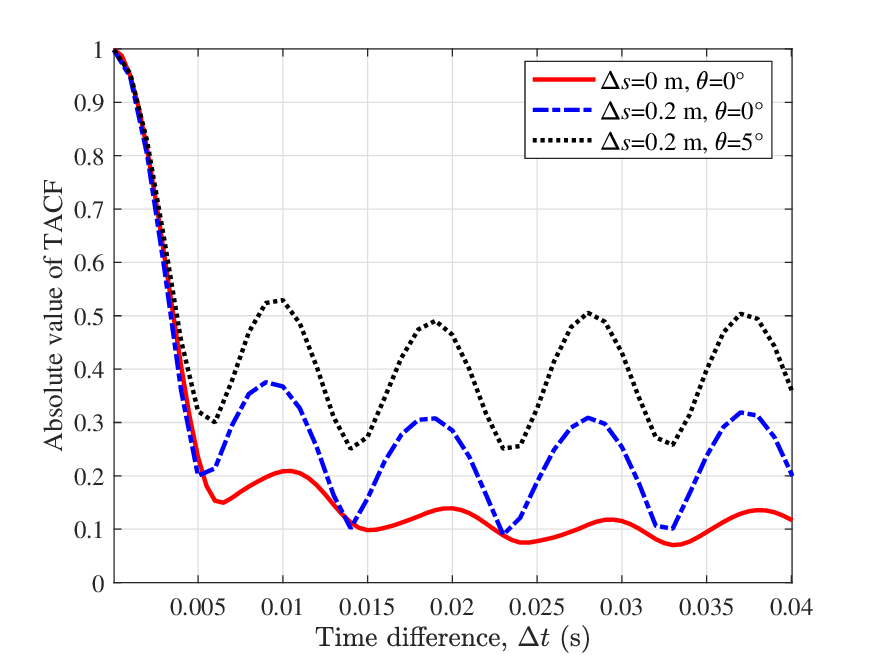}}
			\caption{TACFs  in the cave scenario with different sets of surface roughness and wall tilt (\mbox{$f_{\rm{c}}$ = 2.5 GHz}, \mbox{${v}_{\rm{R}}$ = 5 m/s}, \mbox{${v}_{\rm{T}}$ = 8 m/s}, \mbox{$\alpha _{\rm{T(R)}}=\beta _{\rm{T(R)}} = 0$},\mbox{ $\gamma(t)=\beta(t)=\varphi(t) = 0$}).}
			\label{TACF_RouTilt}
		\end{figure}

		Fig.~\ref{TACF_RouTilt} shows the TACF in the cave scenario with different sets of surface roughness and wall tilt.
		Under the same wall tilt, greater surface roughness results in more pronounced fluctuations in the TACF. This behavior arises because rougher surfaces promote the generation of a larger number of DMCs, thereby intensifying the multipath effect.	
		Additionally, for a given surface roughness, increasing wall tilt improves channel correlation.
		This is because increased wall tilt alters signal reflection paths and phase relationships. Consequently, the dominance of the LoS component is enhanced, leading to higher correlation.
				
		
\begin{figure}[!t]
			\centering	{\includegraphics[width=3.5in]{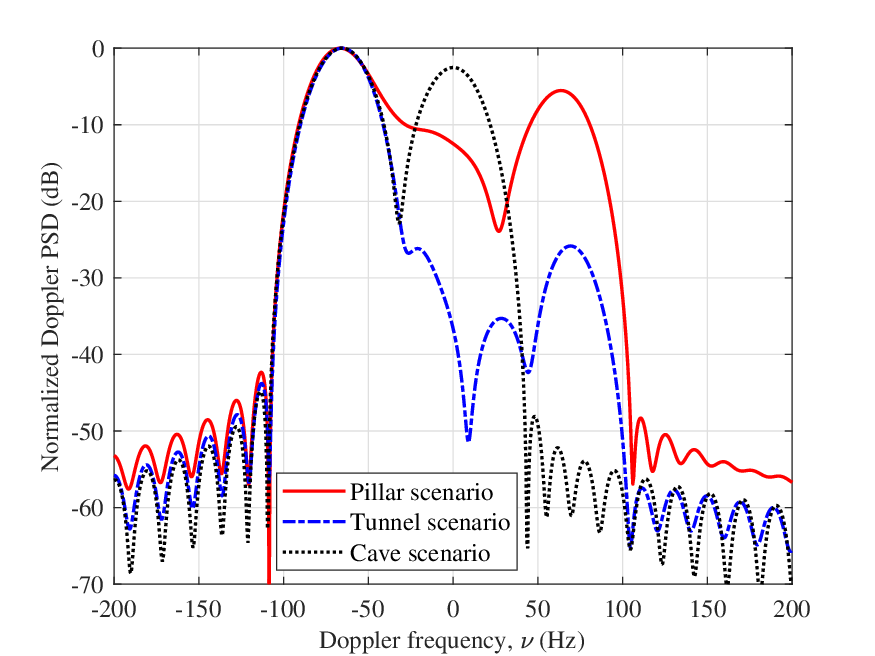}}
			\caption{Doppler PSDs of the proposed channel model in different underground scenarios (\mbox{$f_{\rm{c}}$ = 2.5 GHz}, \mbox{${v}_{\rm{R}}$ = 5 m/s}, \mbox{${v}_{\rm{T}}$ = 8 m/s}, \mbox{$\alpha _{\rm{T(R)}} = \pi/3$}, \mbox{$\beta _{\rm{T(R)}} = \pi/4$}, \mbox{ $\gamma(t)=\beta(t) = \pi/4$}, $\varphi(t)=0$).}
			\label{DopplerPSD}		
		\end{figure}

		Fig.~\ref{DopplerPSD} presents the normalized Doppler PSD in different underground scenarios.
		The tunnel scenario exhibits a narrow and prominent peak,  indicative of limited Doppler spread due to waveguide effects.
		In contrast, the pillar and cave scenarios show multiple distinct peaks, reflecting significant variations in the Doppler shifts among MPCs.
		The Doppler shift depends on the propagation paths and the relative velocity between the Tx/Rx and each MPC. The negative Doppler shifts arise when MPCs propagate in directions opposite to the movement of transceiver.
		Spatially dispersed clusters in pillar and cave scenarios generate both forward- and backward-propagating MPCs, resulting in discrete peaks in both the positive and negative Doppler regions.	
		Together with the results in Fig.~\ref{TACF}, a smaller Doppler spread leads to a larger coherence time, indicating higher temporal correlation in underground scenarios.
				
	
\begin{figure}[!t]
			\centering	{\includegraphics[width=3.5in]{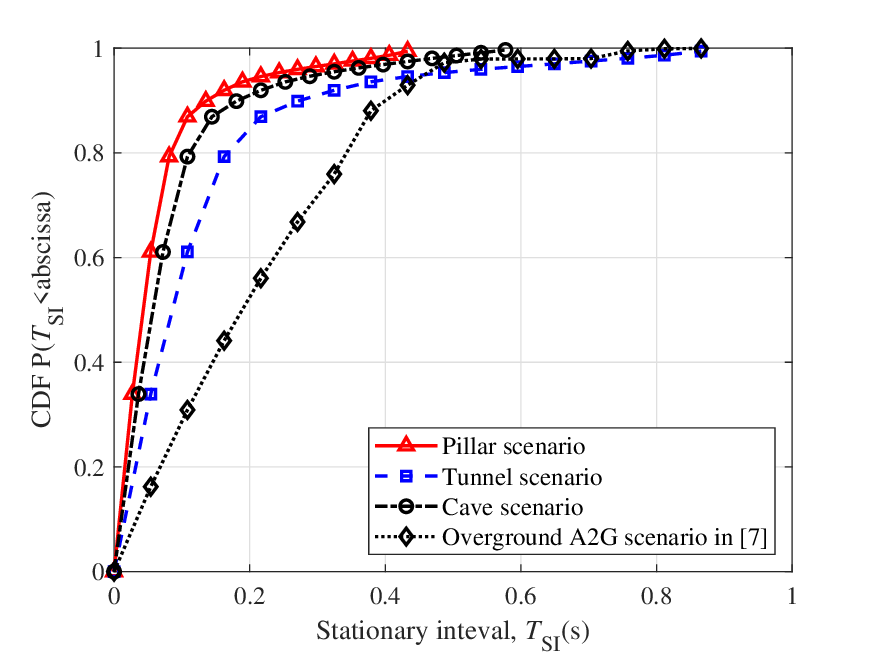}}
			\caption{CDFs of stationary interval for the proposed channel model in different underground scenarios and the channel model in the overground A2G scenario in~\cite{J21_IOT_HChang_UAVGBSM} (\mbox{$f_{\rm{c}}$ = 2.5 GHz}, \mbox{${v}_{\rm{R}}$ = 0 m/s}, \mbox{${v}_{\rm{T}}$ = 5 m/s}, \mbox{$\alpha _{\rm{T(R)}} = \pi/3$}, \mbox{$\beta _{\rm{T(R)}} = \pi/4$}, \mbox{ $\gamma(t)=\beta(t) = \pi/4$}, $\varphi(t)=0$).}
			\label{SI}
		\end{figure}

		Fig.~\ref{SI} shows the CDFs of stationary intervals in different underground scenarios. 
		For comparison, the stationary interval derived from the channel model proposed in \cite{J21_IOT_HChang_UAVGBSM} for the overground A2G scenario is also provided.
		Within underground spaces, the tunnel scenario exhibits the shortest stationary interval, the pillar scenario the largest, and the cave scenario falls in between.
		The stationary interval ordering in Fig.~\ref{SI} is determined by the spatial cluster structure and its evolution mechanism. In the pillar scenario, the wider delay distribution is reshaped rapidly as spatially dispersed clusters enter and leave the VR. In the tunnel scenario, the MPC delays are concentrated within a narrower range, and the delay PSD changes more gradually because the powers of directionally confined clusters evolve smoothly during the birth-death process. In the cave scenario, the coexistence of both mechanisms leads to intermediate behavior. The overground A2G channel model has the longest stationary interval because cluster visibility changes more gradually in the open environment. Thus, Fig.~\ref{RMSDS} characterizes delay dispersion, whereas Fig.~\ref{SI} characterizes the temporal evolution of the delay PSD.
	
\begin{figure}[!t]
			\centering	{\includegraphics[width=3.5in]{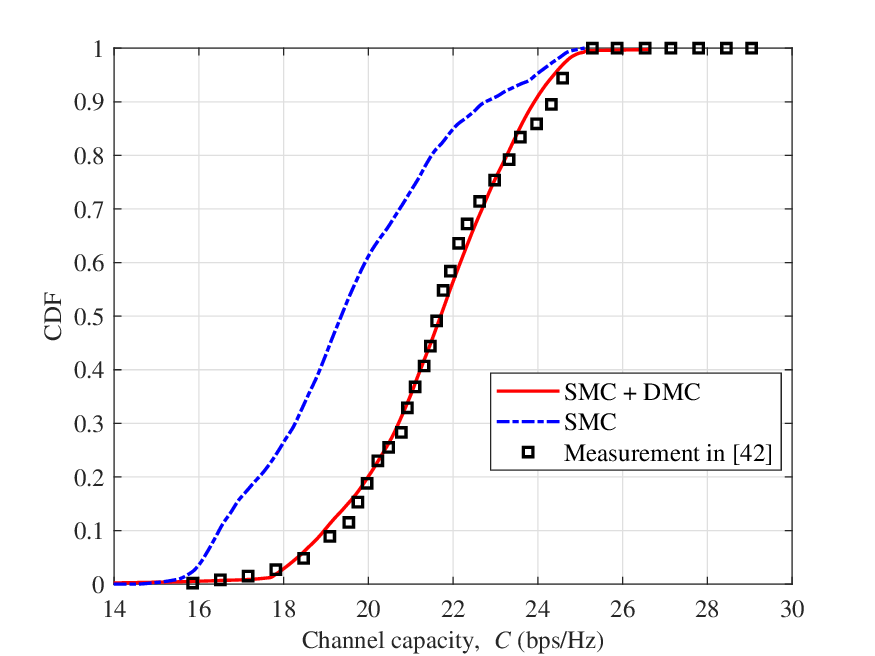}}
			\caption{CDFs of channel capacity in the tunnel scenario from the simulation and the measurement in~\cite{J20_VTM_AAziminejad_TunnelmeaCapacitycom} (\mbox{$f_{\rm{c}}$ = 2.4 GHz}, \mbox{${v}_{\rm{R}}$ = 5 m/s}, \mbox{${v}_{\rm{T}}$ = 8 m/s}, \mbox{$\alpha _{\rm{T(R)}} = \pi/3$}, \mbox{$\beta _{\rm{T(R)}} = \pi/4$}, \mbox{ $\gamma(t)=\beta(t)=\varphi(t)= 0$}).}
			\label{Capacitycom}
		\end{figure}

		Fig.~\ref{Capacitycom} illustrates the CDFs of channel capacity in the tunnel scenario with and without DMCs.
		Moreover, the channel capacity from \cite{J20_VTM_AAziminejad_TunnelmeaCapacitycom} is provided for comparison. In \cite{J20_VTM_AAziminejad_TunnelmeaCapacitycom}, channel measurements were obtained at 2.4~GHz using 2×2~MIMO transceiver antenna arrays in a subway tunnel, under a receiver signal-to-noise ratio of 100~dB.		
		When only SMCs are considered, the channel capacity is underestimated.
		In contrast, the proposed channel model accounting for both SMCs and DMCs can effectively approximate the actual channel capacity.	
		This improvement stems from the ability of model to capture more realistic multipath characteristics. 

	\section{Conclusions}\label{p5}
		In this paper, a general 3D non-stationary MIMO channel model has been proposed for autonomous vehicle communications in underground spaces, applicable to pillar, tunnel, and cave scenarios.
		The channel non-stationarity and key propagation characteristics such as the waveguide effect and near-field/far-field effects have been considered in the proposed channel model.					
		The generation and evolution mechanisms of clusters and DMCs/SMCs have been integrated based on the structural and physical features of underground spaces.
		Furthermore, key statistical properties have been derived and simulated, where the simulated results are in high agreement with the measurements.			
		Results have demonstrated that under the same condition, the channel correlation decreases in sequence across tunnel, cave, and pillar scenarios, which is attributed to increasing structural complexity and multipath randomness.
		Moreover, the incorporation of DMCs has enabled more accurate channel capacity characterization, reduced frequency correlation overestimation, and thereby enhanced the reliability of channel modeling and system design. 
		Compared with autonomous vehicle communication systems in overground spaces, the design of such systems in underground spaces must address the intricate and dynamic multipath effects inherent to these unique environments. 
		Although validated only at 2.5~GHz, the proposed channel model can be extended to other frequencies through measurement-based validation and parameter calibration.


\begin{thebibliography}{00}
		\bibitem{J23_ComST_CXWang_6G}
			C.-X. Wang, \emph{et al}., ``On the road to 6G: Visions, requirements, key technologies and testbeds,'' \emph{IEEE Commun. Surveys Tuts.}, vol.~25, no.~2, pp.~905--974, 2$^{\text{nd}}$ Quart.~2023.
		\bibitem{J22_TVT_CXWang_PervasiveCM1}
			C.-X. Wang, Z. Lv, X. Gao, X.-H. You, Y. Hao, and H. Haas, ``Pervasive wireless channel modeling theory and applications to 6G GBSMs for all frequency bands and all scenarios,'' \emph{IEEE Trans. Veh. Technol.}, vol.~71, no.~9, pp.~9159--9173, Sept.~2022.
		\bibitem{J09_ComST_SYarkan_Underground}
			S. Yarkan, S. Guzelgoz, H. Arslan and R. R. Murphy, ``Underground mine communications: A survey,'' \emph{IEEE Commun. Surveys Tut.}, vol.~11, no.~3, pp.~125--142, 3$^{\text{rd}}$ Quart.~2009.			
		\bibitem{J19_TVT_HQin_UAVundergroundSurvey}
			H. Qin, \emph{et al}., ``Autonomous exploration and mapping system using heterogeneous UAVs and UGVs in GPS-denied environments,'' \emph{IEEE Trans. Veh. Technol.}, vol.~68, no.~2, pp.~1339--1350, Feb.~2019.
		\bibitem{J21_AESM_MDePetrillo_UAVundergroundSurvey}
			M. De Petrillo, J. Beard, Y. Gu, and J. N. Gross, ``Search planning of a UAV/UGV team with localization uncertainty in a subterranean environment,'' \emph{IEEE Aerosp. Electron. Syst. Mag.}, vol.~36, no.~6,  \mbox{pp.~6--16}, Jun.~2021.	
		\bibitem{J20_TVT_ZMa_UAVGBSMA2A}	
			Z. Ma, B. Ai, R. He, G. Wang, Y. Niu, and Z. Zhong, ``A wideband non-stationary air-to-air channel model for UAV communications,'' \emph{IEEE Trans. Veh. Technol.},  vol.~69, no.~2, pp.~1214--1226, Feb.~2020.
		\bibitem{J21_IOT_HChang_UAVGBSM}
			H. Chang, \textit{et al.}, ``A novel non-stationary 6G UAV-to-ground wireless channel model with 3D arbitrary trajectory changes,'' \emph{IEEE Internet Things J.}, vol.~8, no.~12, pp.~9865--9877, Jun.~2021.
		\bibitem{J25_IOT_SYang_UAVwobbling}	
			S. Yang, Z. Zhang, J. Zhang, X. Chu, and J. Zhang, ``Adaptive modulation for wobbling drone air-to-ground links in millimeter-wave bands,'' \emph{IEEE Internet Things J.}, vol.~12, no.~8, pp.~9792--9804, Apr. 2025.		
		\bibitem{J23_TVT_ZHuang_UAVGBSM}
			Z. Huang, L. Bai, M. Sun, and X. Cheng, ``A 3D non-stationarity and consistency model for cooperative multi-vehicle channels,'' \emph{IEEE Trans. Veh. Technol.}, vol.~72, no.~9, pp.~11095--11110, Sept.~2023.
		\bibitem{J25_TWC_BHua_UAVGBSM}
			B. Hua, \textit{et al.}, ``Ultra-wideband nonstationary channel modeling for UAV-to-ground communications,'' \emph{IEEE Trans. Wireless Commun.}, vol.~24, no.~5, pp.~4190--4204, May~2025.
		\bibitem{J24_TCNC_BHe_UAVundergroundSurvey}
			B. He, X. Ji, G. Li, and B. Cheng, ``Key technologies and applications of UAVs in underground space: A review,'' \emph{IEEE Trans. Cognit. Commun. Netw.}, vol.~10, no.~3, pp.~1026--1049, Jun.~2024.	
		\bibitem{J09_TAP_WXu_Surfaceroughness}
			W. Xu, S. A. Zekavat, and H. Tong, ``A novel spatially correlated multiuser MIMO channel modeling: Impact of surface roughness,'' \emph{IEEE Trans. Antennas Propag.}, vol.~57, no.~8, pp.~2429--2438, Aug.~2009.
		\bibitem{J75_TAP_AEmslie_Undergroundtunnel}
			A. Emslie, R. Lagace, and P. Strong, ``Theory of the propagation of UHF radio waves in coal mine tunnels,'' \emph{IEEE Trans. Antennas Propag.}, vol.~23, no.~2, pp.~192--205, Mar.~1975.
		\bibitem{C11_EUCAP_MJacob_Blockageeffect}
			M. Jacob, \emph{et al}., ``A ray tracing based stochastic human blockage model for the IEEE 802.11ad 60 GHz channel model'' in \emph{Proc. EuCAP’11}, Rome, Italy, 2011, pp.~3084--3088.
		\bibitem{J19_CC_MYang_Pillarmea}	
			M. Yang, \emph{et al.}, ``V2V channel characterization and modeling for underground parking garages,'' \emph{China Commun.}, vol.~16, no.~9, \mbox{pp.~93--105}, Sep.~2019.
		\bibitem{J23_TAP_HChang_TunnelPillarGBSM}
			H. Chang, \emph{et al.}, ``A general 3-D nonstationary GBSM for underground vehicular channels,'' \emph{IEEE Trans. Antennas Propag.}, vol.~71, no.~2, pp.~1804--1819, Feb.~2023.
		\bibitem{J03_JSAC_MLienard_Tunnelmea}	
			M. Lienard, P. Degauque, J. Baudet, and D. Degardin, ``Investigation on MIMO channels in subway tunnels,'' \emph{IEEE J. Sel. Areas Commun.}, vol.~21, no.~3, pp.~332--339, Apr.~2003.
		
		\bibitem{J07_TVT_CBrisoRodriguez_UndergroundtunnelBP}	
			C. Briso-Rodriguez, J. M. Cruz, and J. I. Alonso, ``Measurements and modeling of distributed antenna systems in railway tunnels,'' \emph{IEEE Trans. Veh. Technol.}, vol.~56,~pp.~2870--2879, Sept. 2007.
		\bibitem{J19_AWPL_FChallita_UndergroundDMCmea}		
			F. Challita, \emph{et al.}, ``On the contribution of dense multipath components in an intrawagon environment for 5G mmW massive MIMO channels,'' \emph{IEEE Antennas Wireless Propag. Lett.}, vol.~18, pp.~2483--2487, Dec.~2019.
		\bibitem{C20_EuCAP_SJiang_UndergroundDMCmea}	
			S. Jiang, X. Zhang, W. Wang, M. Yang, and R. He, ``Measurement and diffuse multipath analysis of V2V propagation channel at 5.9 GHz in tunnel area,'' in \emph{Proc. EuCAP’20}, Copenhagen, Denmark, 2020, pp.~1--5.
		\bibitem{J22_APM_ASaleem_UndergroundDMCmea}
			A. Saleem, H. Cui, Y. He, and A. Boag, ``Channel propagation characteristics for massive multiple-input/multiple-output systems in a tunnel environment,'' \emph{IEEE Antennas Propag. Mag.}, vol.~64, no.~3, pp.~126--142, Jun.~2022.	
		\bibitem{J14_TAP_MDBedford_Cave}
			M. D. Bedford and G. A. Kennedy, ``Modeling microwave propagation in natural caves passages,'' \emph{IEEE Trans. Antennas Propag.}, vol.~62, no.~12, pp.~6463--6471, Dec.~2014.	
		\bibitem{J10_TCOM_ZSun_Undergroundtunnelpillar}
			Z. Sun and I. F. Akyildiz, ``Channel modeling and analysis for wireless networks in underground mines and road tunnels,'' \emph{IEEE Trans. Commun.}, vol.~58, no.~6, pp.~1758--1768, Jun.~2010.
		\bibitem{J19_TAP_XZhang_Undergroundtunnel}
			X. Zhang and C. D. Sarris, ``Statistical modeling of electromagnetic wave propagation in tunnels with rough walls using the vector parabolic equation method,'' \emph{IEEE Trans. Antennas Propag.}, vol.~67, no.~4, pp.~2645--2654, Apr.~2019.	
		\bibitem{C21_WCNC_MWu_Undergroundpillarroughness}
			M. Wu, Y. Ma, and Z. Zhu, “Ray tracing-based wireless channel modeling in room-and-pillar mines,'' in \emph{Proc. WCNC’21}, Nanjing, China, 2021, pp.~1--6.	
		\bibitem{J25_SCIS_CXWang_Enhanced6G}
			C.-X. Wang, \textit{et al.}, ``An enhanced 6G pervasive channel model towards standardization,'' \emph{Sci. China Inf. Sci.}, vol.~68, no.~6, Jun.~2025, doi: 10.1360/SSI-2025-0029.	
		\bibitem{J16_TVT_NAvazov_TunnelGBSM}
			N. Avazov and M. Pätzold, ``A novel wideband MIMO car-to-car channel model based on a geometrical semi-circular tunnel scattering model,'' \emph{IEEE Trans. Veh. Technol.}, vol.~65, no.~3, pp.~1070--1082, Mar.~2016.
		\bibitem{J20_TITS_YLiu_TunnelGBSM}
			Y. Liu, \textit{et al.}, ``3D non-stationary wideband tunnel channel models for 5G high-speed train wireless communications,'' \emph{IEEE Trans. Intell. Transp. Syst.}, vol.~21, no.~1, pp.~259--272, Jan.~2020.				
		\bibitem{C23_ICCC_XJi_SurfaceroughnessBlockageGBSM}
			X. Ji, C.-X. Wang and H. Chang, ``A novel MIMO channel model for underground mine communications,'' in \emph{Proc. ICCC’23}, Dalian, China, 2023, pp.~1--5.			
		\bibitem{C15_EuCAP_UTVirk_DMCgbsm}	
			U. T. Virk, K. Haneda, and J.-F. Wagen, ``Dense multipath components add-on for COST 2100 channel model,'' in \emph{Proc. EuCAP’15}, Lisbon, Portugal, 2015, pp.~1--5.
		\bibitem{C21_PIMRC_YLi_DMCgbsm}		
			Y. Li, C.-X. Wang, and Y. Liu, ``A 3D non-stationary geometry-based stochastic model for industrial automation wireless communication systems,'' in \emph{Proc. IEEE PIMRC’21}, Helsinki, Finland, 2021, \mbox{pp.~916--921}.
		\bibitem{J25_TVT_LZhang_DMCgbsm}
			L. Zhang, \emph{et al.}, ``A novel geometry-based stochastic model for indoor scenarios incorporating dense multipath components towards standardization,'' \emph{IEEE Trans. Veh. Technol.},  vol.~74, no.~7, pp.~10927--10942, Jul.~2025.
		\bibitem{J10_AWPL_JPoutanen_DMCmedel}
			J. Poutanen, J. Salmi, K. Haneda, V.-M. Kolmonen, F. Tufvesson, and P. Vainikainen, ``Propagation characteristics of dense multipath components,'' \emph{IEEE Antennas Wireless Propag. Lett.}, vol.~8, \mbox{pp.~791--794}, Aug.~2010.	
		\bibitem{3GPP38.901}
			3GPP, ``Study on channel model for frequencies from 0.5 to 100 GHz,'' 3GPP TR 38.901, Tech.~Rep., V17.1.0, Dec.~2023.	
		\bibitem{C11_VTC_KGuan_UndergroundtunnelBP}	
			K. Guan, Z. Zhong, B. Ai, and C. Briso-Rodriguez, ``Propagation mechanism analysis before the break point inside tunnels,'' in \emph{Proc. VTC Fall'11}, San Francisco, CA, USA, 2011, pp.~1--5.
		\bibitem{J07_TAP_MRak_TunnelCave}
			M. Rak and P. Pechac, ``UHF propagation in caves and subterranean galleries,'' \emph{IEEE Trans. Antennas Propag.}, vol.~55, no.~4, pp.~1134--1138, Apr.~2007.
		\bibitem{J19_AWPL_JLi_IndoorDMC}
			J. Li, B. Ai, R. He, M. Yang, and Z. Zhong, ``On modeling of dense multipath component for indoor massive MIMO channels,'' \emph{IEEE Antennas Wireless Propag. Lett.}, vol.~18, no.~3, pp.~526--530, Mar.~2019.
		\bibitem{J17_TAP_CZhou_Surfaceroughness}
			C. Zhou, ``Ray tracing and modal methods for modeling radio propagation in tunnels with rough walls,'' \emph{IEEE Trans. Antennas Propag.}, vol.~65, no.~5, pp.~2624--2634, May~2017.	
		\bibitem{J23_TCOM_CXWang_PervasiveCM2}
			C.-X. Wang, Z. Lv, Y. Chen, and H. Haas, ``A complete study of space-time-frequency statistical properties of the 6G pervasive channel model,'' \emph{IEEE Trans. Commun.}, vol.~71, no.~12, pp.~7273--7287, Dec.~2023.
		\bibitem{J14_TCOM_AEForooshani_Tunnelangular}
			A. E. Forooshani, S. Noghanian, and D. G. Michelson, ``Characterization of angular spread in underground tunnels based on the multimode waveguide model,'' \emph{IEEE Trans. Commun.}, vol.~62, no.~11, \mbox{pp.~4126--4133}, Nov.~2014.
		\bibitem{J08_TAP_MBoutin_TunnelmeaDScom}
			M. Boutin, A. Benzakour, C. L. Despins, and S. Affes, ``Radio wave characterization and modeling in underground mine tunnels,'' \emph{IEEE Trans. Antennas Propag.}, vol.~56, no.~2, pp.~540--549, Feb.~2008.
		\bibitem{J20_VTM_AAziminejad_TunnelmeaCapacitycom}
			A. Aziminejad and Y. He, ``Radio communication in curved tunnels: MIMO channel capacity for rail transit applications,'' \emph{IEEE  Veh. Technol. Mag.}, vol.~15, no.~1, pp.~99--106, Mar.~2020.
		
	
	\end{thebibliography}
\end{document}